\documentclass[mypaper]{myaa}
\usepackage{natbib}
\bibpunct{(}{)}{;}{a}{}{,} 
\usepackage{hyperref}
 \usepackage{aalongtable}
\usepackage{graphicx}
\usepackage{txfonts}
 \usepackage{balance}
\usepackage{latexsym}
\newcommand{\nc}{\newcommand}  
\nc{\teff}{$T_{\rm eff}$\,}  
\nc{\teffns}{$T_{\rm eff}$}  
\nc{\glog}{log\,$g$\,}  
\nc{\kms}{\,${\rm km\,s}^{-1}$\,}  
\nc{\mic}{$\xi_{\rm t}$\,}
\nc{\nod}{--}
\idline{0}{1}
\begin{document}

\title{{First stars VII.  
Lithium in extremely metal poor dwarfs}
\subtitle{}
\thanks {Based on observations made with the ESO Very Large Telescope 
at Paranal Observatory, Chile (Large Programme ``First Stars'', 
ID 165.N-0276(A); P.I. R. Cayrel).}
}

\author{
P. Bonifacio\inst {1,2,3} \and
P. Molaro\inst {2,3} \and
T. Sivarani \inst{4} \and
R. Cayrel\inst {2} \and
M. Spite\inst {2} \and
F. Spite \inst {2} \and
B. Plez\inst {5} \and
J.~Andersen\inst {6,7} \and
B.~Barbuy\inst {8} \and
T.~C.~Beers\inst {4} \and
E. Depagne \inst{9} \and
V. Hill\inst {2} \and 
P. Fran\c cois  \inst{2} \and
B.~Nordstr\"om \inst {6} \and 
F. Primas\inst {10}
} 

\institute{ 
CIFIST Marie Curie Excellence Team
\and
Observatoire de Paris, GEPI,
             F-92195 Meudon Cedex, France\\
\email{Piercarlo.Bonifacio@obspm.fr, Roger.Cayrel@obspm.fr}\\
   \email {Monique.Spite@obspm.fr, Francois.Spite@obspm.fr}\\
   \email {Vanessa.Hill@obspm.fr,Patrick.Francois@obspm.fr }
         \and
    Istituto Nazionale di Astrofisica - Osservatorio Astronomico di
Trieste,
    Via Tiepolo 11, I-34131
             Trieste, Italy\\
   \email {molaro@ts.astro.it}
\and
 Department of Physics \& Astronomy and JINA: Joint Institute for Nuclear
Astrophysics, Michigan State University,
             East Lansing, MI 48824, USA\\
   \email {thirupati@pa.msu.edu,beers@pa.msu.edu}
         \and
GRAAL, Universit\'e de Montpellier II, F-34095 
Montpellier
             Cedex 05, France\\
   \email {Bertrand.Plez@graal.univ-montp2.fr}
         \and
         The Niels Bohr Institute, Astronomy, Juliane Maries Vej 30,
         DK-2100 Copenhagen, Denmark\\
   \email {ja@astro.ku.dk, birgitta@astro.ku.dk}
  \and
     Nordic Optical Telescope, Apartado 474, E-38700 Santa Cruz de 
     La Palma, Spain\\
   \email {ja@not.iac.es}
         \and
    Universidade de Sao Paulo, Departamento de Astronomia,
Rua do Matao 1226, 05508-900 Sao Paulo, Brazil\\
   \email {barbuy@astro.iag.usp.br}
               \and
European Southern Observatory, Casilla 19001, Santiago, Chile\\
\email{edepagne@eso.org}
        \and 
         European Southern Observatory (ESO),
         Karl-Schwarschild-Str. 2, D-85749 Garching b. M\"unchen, 
Germany\\
   \email {fprimas@eso.org}
}
\authorrunning{Bonifacio et al.}
\titlerunning{First stars VII. Lithium in EMP dwarfs}
\offprints{P. Bonifacio}
\date{Received xxx; Accepted xxx}
\abstract
{The primordial lithium abundance is a key prediction of models 
of big bang nucleosynthesis, and its abundance in metal-poor dwarfs 
(the {\em Spite plateau}) is an important, independent observational 
constraint on such models.}
{This study aims to determine the level and constancy of the Spite 
plateau as definitively as possible from homogeneous high-quality 
VLT-UVES spectra of 19 of the most metal-poor dwarf stars known.}
{Our high-resolution ($R\sim 43000$), high S/N spectra are analysed 
with {\tt OSMARCS} \rm 1D LTE model atmospheres and {\tt turbospectrum} 
\rm synthetic spectra to determine effective temperatures, surface 
gravities, and metallicities, as well as Li abundances for our stars.}
{Eliminating a cool subgiant and a spectroscopic binary, we find 8 stars 
to have $-3.5 < $[Fe/H]$ < -3.0$ and 9 stars with $-3.0 < $[Fe/H]$ < -2.5$.
Our best value for the mean level of the plateau is A(Li) $=2.10\pm 0.09$. 
 The scatter around the mean is entirely explained by our estimate of the 
observational error and does not allow for any intrinsic scatter in the 
Li abundances. In addition, we conclude that 
a systematic error of the order of 200 K in any of the current 
temperature scales remains possible.  The iron excitation equilibria
in our stars support our adopted temperature scale, which is based on a 
fit to wings of the H$\alpha$ line, and disfavour hotter scales, which 
would lead to a higher Li abundance, but fail to achieve excitation 
equilibrium for iron.} 
{We confirm the previously noted discrepancy between the Li abundance
measured in extremely metal-poor turnoff stars and the primordial Li 
abundance predicted by standard Big-Bang nucleosynthesis models adopting 
the baryonic density inferred from WMAP.  We discuss recent work explaining 
the discrepancy in terms of diffusion and find that uncertain temperature 
scales remain a major question.}
\keywords{Nucleosynthesis -- Stars: abundances --  
	  -- Galaxy: Halo -- Galaxy: abundances -- Cosmology: observations}
\maketitle

\section{Introduction}

Metal-poor stars in the Galactic halo provide a fossil record of the chemical
composition of the early Galaxy. Although this is true for most elements in both
dwarfs and unmixed giants, in the case of the fragile 
element Li, it applies
only to dwarf stars. Li is easily diluted in the atmospheres of cool giant stars
when material that has experienced temperatures in excess of $2.5\times 10^6$ K,
and is therefore Li-depleted, is mixed into the outer atmosphere of the star.
Although, in principle, it would be possible to measure Li in early giants and
correct the value using the dilution factor derived from models, in practice the
models are not yet realistic enough to exploit this strategy. As a result, 
to study Li in the early Galaxy, one has to study dwarfs or early
subgiants. Because of their low intrinsic luminosity, it is more difficult to
observe large samples of metal-poor halo dwarfs, but the key role of Li in
constraining the nature of the early Universe amply justifies the effort.

Spite \& Spite \citeyearpar{spite82,ss82A} first demonstrated that metal-poor
dwarfs share the same measured Li abundance regardless of temperature and
metallicity in the range $\rm 5700~K \la T_{eff} \la 6250$~K and $\rm -2.4 \le
[Fe/H] \le -1.4 $. This behaviour is distinctively different from that of other
elements, whose abundances generally drop with declining metallicity. They
interpreted this {\em plateau} (hereafter {\em Spite plateau}) as a signature of
the nucleosynthesis in the hot and dense early phase of the Universe. 

This observation confirmed the theoretical prediction of \citet{wagoner}, who
computed the nucleosynthesis in material with temperatures above $10^{9}$ K on
the short timescales (10s-10$^3$s) appropriate to the first phases of a hot and
dense expanding Universe (the Big Bang), and showed that a non-negligible amount of
Li could be produced in this manner. The most straightforward interpretation of
the Cosmic Microwave Background (CMB) and the Hubble Law requires that the Universe
has indeed passed through such a hot and dense phase, and the computations of
\citet{wagoner} demonstrated that the primordial abundances of $^7$Li, $^4$He,
$^3$He, and D depend on the {\em a-priori} unknown density of baryons. 

In the original interpretation of the {\em Spite plateau}, the Li abundance in
metal-poor dwarfs is virtually equal to the primordial value preserved in the
atmospheres of these stars due to their shallow convection zones. The mixing of
primordial matter with the ejecta of SN II, where Li has been burned, may
slightly lower the Li abundance with respect to the primordial one. Note that
this contrasts with the situation in the Sun, where the photospheric Li is
depleted by two orders of magnitude with  respect to the initial (meteoritic) 
value. 

Since its discovery, the {\em Spite plateau} has been subject to numerous
investigations, increasing the number of stars with Li measurements and
extending the sample to include ever lower metallicities. Several recent studies
have shown that the {\em Spite plateau} exhibits very little, if any, dispersion
\citep{BM97,ryan,melendez,CP05}. 

There are, however, several concerns regarding the identification of the Li
abundance of the {\em Spite plateau} with the primordial value. The most serious
challenge comes from the determination of the baryonic density obtained from the
spectrum of fluctuations in the CMB observed by the WMAP satellite
\citep{spergel,spergel06}. The baryon-to-photon ratio, $\eta$, once considered a
free parameter in standard Big Bang nucleosynthesis (SBBN), is constrained by
these observations to be $\eta = 6.11\pm 0.22\times 10^{-10}$. When inserted
into SBBN computations, this value implies a primordial Li abundance of A(Li)
\footnote{A(Li) = log[N(Li)/N(H)] +12} = 2.64. The highest values claimed for
the {\em Spite plateau} \citep{B02,melendez} are about 0.3 dex lower; many other
recent claims are lower still. If one accepts the WMAP determination of $\eta$,
there are three possibilities: Either the SBBN computations are wrong, the Li
seen in halo dwarfs does not represent the primordial value, or the current
temperature scales are far too cool and the main-sequence turnoff of metal-poor
stars lies at \teff $\sim 7300$~K (\citealt{RM06}).
 
Another challenge to the primordial interpretation of the {\em Spite plateau}
arises from claims that a  slope in Li abundance vs. [Fe/H] exists, in the 
range 0.1--0.2 dex/dex \citep{ryan96,ryan,boesgaard,asplund}.
However, other investigators employing different temperature scales than these
authors failed to detect any slope \citep{spite96,BM97,B2002,melendez} or found
only a very shallow one \citep{CP05}. If one interprets the slope as evidence
for Li production in the early Galaxy, then the primordial Li value should be
obtained by extrapolating the slope down to the lowest metallicities,
exacerbating the discrepancy with the primordial Li implied by the baryonic
density determined by WMAP. 

Our Large Programme~  ``First Stars'' was designed, {\em inter alia}, 
to significantly enlarge 
the sample of extremely metal-poor main-sequence turnoff stars with available
high-resolution spectroscopy, to shed new light on the behaviour of the
{\em Spite plateau} at the lowest metallicities. Prior to our observations, only
ten dwarfs with [Fe/H] $\le -3$ had measured Li abundances (BD --13 3442,
BS~16968-061, CS~22884-108, CS~29527-015, CD --24 17504, CD --33 01173, 
G~064-012, G~064-037, LP~815-43, and LP~831-70), as well as three binary stars
(CS~22873-139, CS~22876-032, and HE~1353-2735); the present observations add to
this sample another seven new extremely metal-poor stars  for which iron
and Li abundances, based on high-resolution analysis, are
presented here for the first time.

\begin{figure}
\resizebox{\hsize}{!}{\includegraphics[clip=true]{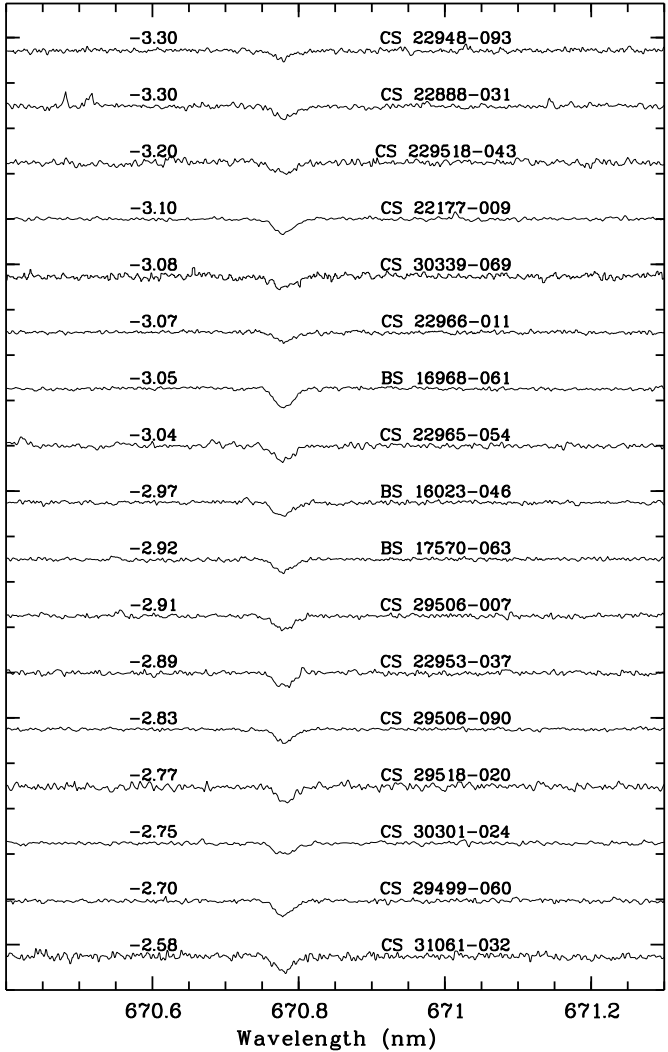}}
\caption{The Li~I doublet in the 17 single main-sequence turnoff (TO) 
stars of our sample.
Multiple spectra of each star have been coadded and normalized; for clarity,
they are  shifted vertically by arbitrary amounts, one tick on the vertical 
axis corresponding to 0.2 in residual intensity. Each spectrum is labelled 
with the name and [Fe/H] of the star.  The stars are shown in order of 
increasing [Fe/H] from top to bottom.}
\label{pl_obs}
\end{figure}

\section{Observations and data reduction}

Our spectroscopic data were obtained as a part of the ESO Large Programme
``First Stars''; the log of observations is given in
Table \ref{logobs}. We used the VLT-Kuyen 8.2m telescope and the UVES
spectrograph \citep{dekker} in two non-standard settings, both with
dichroic \# 1: 396+573, 396+850; the numbers are the central wavelength in nm in
the blue and red arms, respectively. The central wavelengths in the red arm were
chosen in such a way that both settings would cover the Li doublet, essentially
doubling the number of Li spectra for each star. 

Most of the observations
were made with a projected slit width of $1''$, yielding a resolving power
of $R = 43000$. Equivalent-width measurements for unblended lines were
accomplished by fitting Gaussian profiles, using the genetic algorithm code
described in \citet{francois}. The last column of Table \ref{tabli} lists
the S/N ratio per pixel near the Li line.

\section{Atmospheric parameters\label{atm}}

Our analysis used OSMARCS model atmospheres
\citep{Gus75,Ple92,Edv93,Asp97,GEE03} and the {\tt turbospectrum} spectral
synthesis code \citep{alvarez_plez}. Effective temperatures, \teff, for our
programme stars were determined using the wings of H$\alpha$, which is a very
good temperature indicator \citep{cayrel88,fuhrmann,vantveer,barklem02}. We did
not use other Balmer lines because their profiles are sensitive to the
treatment of convection \citep{fuhrmann}. Adopting the broadening theory of
\citet{barklem}, we performed a $\chi^2$ fit of the computed profiles to the
observed spectra. For the H$\alpha$ fitting we assumed log~g = 4.0 for all
stars.

The derived effective temperatures are about 150~K cooler than obtained
previously, using the \citet{vcs} theory (results presented at the IAU General
Assembly in 2003; see Bonifacio et al. 2003). The error in our effective
temperatures is only very weakly dependent on the photon noise, given the large
number of pixels in our spectra ($\sim 5\times10^4$) that define the H$\alpha$
wings. A Monte Carlo simulation showed that at S/N=50/1 the error is of only 12~K. 

A more important source of error is the slight gravity dependence of H$\alpha$,
which is about $-50$~K for a change of +0.25 dex in log~g. In principle, it is
possible to re-determine \teff after log~g has been determined from the iron
ionization equilibrium (see below) and iterate the process. The worst possible
case is CS~22888-031 (\teff = 6151~K, log~g = 5.00), which, after the
iterations, would end up at a 
\teff 150~K lower and a log~g 0.25 dex lower.
In
practice, we feel that our gravities (based on at most four \ion{Fe}{ii}
lines) are too inaccurate to use this approach. 
 Although internally
self-consistent, this method would have introduced more scatter in \teff than 
is implied by our assumption of equal gravity for all stars (for the purpose 
of the \teff determination).
 
A further source of uncertainty is residual curvature and fringing in the
echelle orders and uncertainties in the order merging. These effects are not
easily modelled; to obtain a crude estimate of the associated errors, we
used our observed spectra as templates and introduced the same ``wiggles'' on a
synthetic spectrum. We then performed a model fit to this simulated spectrum and
recovered effective temperatures, which could differ up to about 40~K from the
effective temperature of the model used to compute the synthetic spectrum.

Another source of error is the finite pixel size in our spectra. If we rebin a
synthetic spectrum (computed at a resolution $R = 500,000$) to the same binning as
our observed spectra and fit it again, we recover an effective temperature that
may differ by up to 50~K from the input one, the average error being of about
20~K. By summing these errors linearly (we consider these errors as systematic),
we estimate a total error on \teff ~ of about 130~K. Although this estimate has
been obtained in a somewhat crude manner, we believe there is little prospect
for refining it. Errors as small as 50~K are clearly not realistic, while errors as large as 200~K appear unlikely. For the rest of our discussion it would
make little difference if the error is 110~K or 150~K. Note that this estimate
does {\em not} include any systematic errors in the line broadening theory; as
implied above, this may be in the range 150-200~K. The determination of
temperatures by colours is discussed below (see Sect. \ref{TEFF}).

For each star, we initially adopted a surface gravity of log~g = 4.0 and used
the metallicity derived in \citet{BIAU}. With these parameters and the
H$\alpha$-based \teff, we interpolated a model atmosphere within a grid of
OSMARCS models that was specifically computed for our
application, over the metallicity range $\rm -4\le[Fe/H]\le-2 $, with abundances
of the $\alpha$ elements enhanced by 0.4 dex. Using the equivalent widths of the lines listed in Tables \ref{fe_1}, \ref{fe_2}, \ref{fe_3}, and \ref{fe_4}, 
 where the adopted log gf values and their references are also listed, 
we determined the abundance of 
\ion{Fe}{i} and \ion{Fe}{ii}. The microturbulent velocity was adjusted 
by requiring that strong lines and weak lines yield the same abundance. The
process was iterated by adjusting the gravity until an iron ionization
equilibrium was achieved to within 0.05 dex. The final derived atmospheric
parameters are listed in Table \ref{tabli}. The line-to-line scatter for
\ion{Fe}{i} was typically in the range 0.10 to 0.15 dex.
As the reference solar iron abundance,
we adopted A(Fe)$_{\sun}$= 7.51, which is
the value determined by \citet{anstee} and which 
coincides with the meteoritic value.
The [Fe/H] given in Table \ref{tabli} is the mean \ion{Fe}{i}
abundance.

\section{Lithium abundances\label{abundances}}

The equivalent width (EW) of the Li I doublet for each star was measured by fitting a
synthetic profile, following the procedures discussed by \citet{B02}. The EW
errors listed in Table \ref{tabli} were estimated using Monte Carlo simulations,
in which Poisson noise was added to a synthetic spectrum to reach the
same S/N ratio as the observed spectrum. However, the Poisson noise is not the
only source of error in the EWs. At the wavelength of the Li doublet, CCD
detectors show effects of ``fringing'', which is never totally removed by
flat-fielding. Residual fringing is always present at the level of a few per
cent, and may introduce even larger errors than Poisson noise.

For our data, we estimated this effect by comparing results for the same star as
obtained from spectra in the 573 nm setting (where the Li doublet falls on the
EEV CCD) and in the 850 nm setting (where the Li doublet falls on the
deep-deletion MIT CCD), which show different fringing patterns. We also compared
measurements of the same star on different dates, when the Li line falls on a
different portion of the fringing pattern. We conclude that residual fringing
may introduce an error of about 0.1 pm, which dominates over Poisson noise. An
error of 0.1 pm in the EW results in an error of 0.03 dex in the derived lithium
abundance.

We used {\tt turbospectrum} and the model atmosphere with parameters determined
in Sect. \ref{atm} to iteratively compute synthetic spectra of the doublet
until the synthetic EW matched the measured EW to better than 1\%. 
 The adopted atomic data for the Li doublet is the same as that
used by \citet{asplund}, thus taking into account the hyperfine structure
of the lines and also the isotopic components, for an assumed solar
isotopic ratio. Note that changes in the isotopic ratio or even neglect 
of the isotopic structure have no effect on the Li abundance derived from these weak lines. 
The various
steps of the iteration provide a curve of growth for the Li doublet, which we
used to determine the error in A(Li) arising from the error in EW. For each star
we also determined A(Li) from models with effective temperatures set to $\pm
130$ K with respect to the adopted temperature, which allowed us to determine
the error in A(Li) due to uncertainty in effective temperature; this amounts to
0.09 dex. The errors arising from reasonable uncertainties in surface gravity
and microturbulent velocity are less than 0.01 dex, and can be ignored. The
total error in A(Li) is determined by summing the errors from EW and \teff in
quadrature. Due to the high quality of our data, the final error is dominated by
the uncertainty in \teff.
In the present analysis we have adopted 1D model atmospheres, which ignore the
effects of stellar granulation. Previous computations for the Sun
\citep{Kisel97,Kisel98} and metal-poor stars \citep{CS00,asp03} suggest that
these are not important for the Li doublet. 

\section{Analysis}

Our full sample consists of 19 stars from the HK objective-prism survey of 
Beers and collaborators \citep{beers85,beers92,beers99}, all of which had 
been classified as TO
stars on the basis of medium-resolution spectra and photometry. The
original sample consisted of 27 stars, but we have excluded
 two known spectroscopic binaries, three carbon-rich stars, one Halo
Blue Straggler, and two Horizontal-Branch (HB) stars, from the present
discussion. The carbon-rich stars are
the topic of another paper in this series \citep{siva06}, and the binaries 
will be the subject of yet another paper. 

We did not detect any Li features in either the Halo Blue Straggler or the two
HB stars. This non-detection is consistent with our current understanding of
both types of stars. In HB stars the Li, having already been diluted and
partially destroyed during the star's red giant-branch phase, is fully destroyed
during the He-flash that brings the star onto the Zero Age Horizontal Branch
(ZAHB). The study of Li in high-velocity A- and F-type stars by \citet{glaspey}
clearly indicates that Li is strongly depleted in all Blue Straggler stars; Ryan
et al. \citeyearpar{ryan01,ryan02} explain these depletions as the result of
mass-transfer from a former companion. 

\begin{figure}
\resizebox{\hsize}{!}{\includegraphics[clip=true]{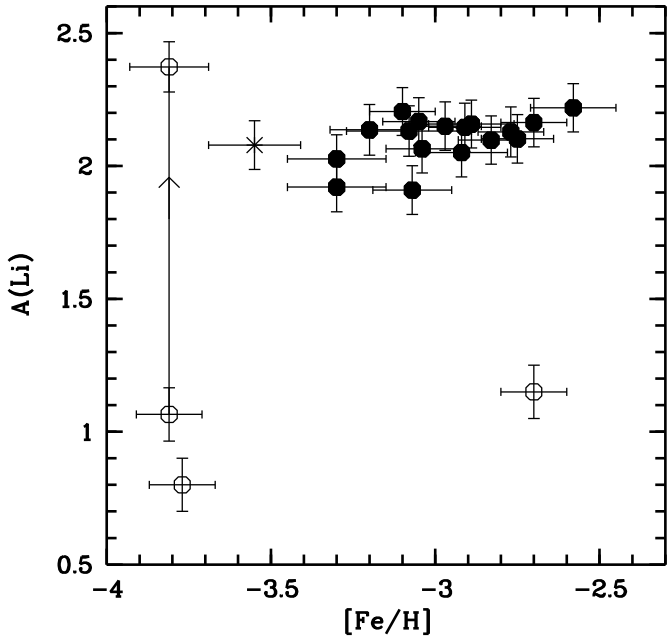}}
\caption{Derived Li abundances, corrected for standard depletion and
NLTE effects, vs. [Fe/H]. The open symbols indicate the cool subgiant 
\object{BS~16076-006} and the two giants/HB
stars of similar \teff, \object{CS~22896-154} and
\object{BS~16467-062} (Cayrel et al. 2004); 
the observed A(Li) and the value corrected for standard depletion
for \object{BS~16076-006} are connected by an arrow;
the asterisk is \object{CS~29527-015}, which 
is a double-lined spectroscopic binary.}
\label{fig_fehli}
\end{figure}

In the remaining sample, \object{BS~16076-006} turned out to be a cool subgiant;
its Li abundance is A(Li)=1.13, considerably lower than in the other stars where
detectable Li was expected. We interpret this as an effect of dilution, as
predicted by standard models. It is interesting to note that an order of
magnitude of the effect in subgiants may be estimated from the Li depletion
isochrones of \citet{deli}. Once corrected for this depletion, as well as for
expected NLTE effects, the pristine value of Li in this subgiant is A(Li) = 2.43
(see Fig. \ref{fig_fehli}), which is above the range spanned by the Li
abundances in the TO stars. From the paper of \citet{ryan98}, an independent
evaluation of the dilution of Li in halo subgiants may be also found; it leads
to a similar result. However, note that the error on the depletion corrections
for subgiants and giants are probably larger than those for dwarfs. Therefore
the above value should be used with some caution.

From spectra taken over several runs, we find \object{CS~29527-015} to be a
double-lined spectroscopic binary (SB2). On our spectra, the measured EW of the
Li doublet varies from 1.0 pm to 1.8 pm, far more than our expected
observational error. This presumably reconciles the non-detection of Li in this
star by \citet{thorburn} and \citet{norris97} with the Li detection by
\citet{spite_natal}; continuum light from the companion star could easily 
fill in the weak Li doublet (see the analysis of the SB2 CS~22876-032 by
\citealt{norris00}).

\begin{table*}
\caption{Atmospheric parameters and Li abundances}
\label{tabli}
\begin{tabular}{rrrrrrrrrrrrr}
\hline
\hline
Star & V   & K   &\teff & \glog &  \mic & [Fe/H] & EW & $\sigma_{EW}$ &A(Li) &A(Li)$_c$& $\mathrm \sigma_{A(Li)}$& S/N\\
     & mag & mag & K    & cgs   & \kms  & dex    & pm & pm            & dex  & dex     & dex & @670nm\\
\hline
\hline
 \object{ BS~16023-046  } & 14.17 & 12.93 & 6364 & 4.50 & 1.3 & -2.97 & 1.93 & 0.060 & 2.18 & 2.15 & 0.09& 186 \\
 \object{ BS~16076-006  } & 13.44 & 11.69 & 5199 & 3.00 & 1.4 & -3.81 & 1.29 & 0.080 & 1.07 & 2.37 & 0.09& 224\\
 \object{ BS~16968-061  } & 13.26 & 11.89 & 6035 & 3.75 & 1.5 & -3.05 & 2.79 & 0.032 & 2.12 & 2.17 & 0.09& 289 \\
 \object{ BS~17570-063  } & 14.51 & 13.19 & 6242 & 4.75 & 0.5 & -2.92 & 1.76 & 0.040 & 2.05 & 2.05 & 0.09& 200 \\
 \object{ CS~22177-009  } & 14.27 & 13.03 & 6257 & 4.50 & 1.2 & -3.10 & 2.42 & 0.030 & 2.21 & 2.20 & 0.09& 178\\
 \object{ CS~22888-031  } & 14.90 & 13.58 & 6151 & 5.00 & 0.5 & -3.30 & 1.87 & 0.040 & 2.01 & 2.03 & 0.09& 157 \\ 
 \object{ CS~22948-093  } & 15.18 & 14.01 & 6356 & 4.25 & 1.2 & -3.30 & 1.19 & 0.060 & 1.94 & 1.92 & 0.09& 181\\ 
 \object{ CS~22953-037  } & 13.64 & 12.43 & 6364 & 4.25 & 1.4 & -2.89 & 1.95 & 0.030 & 2.16 & 2.16 & 0.09& 284 \\
 \object{ CS~22965-054  } & 15.10 & 13.45 & 6089 & 3.75 & 1.4 & -3.04 & 2.21 & 0.060 & 2.03 & 2.06 & 0.09& 150 \\
 \object{ CS~22966-011  } & 14.55 & 13.28 & 6204 & 4.75 & 1.1 & -3.07 & 1.37 & 0.050 & 1.90 & 1.91 & 0.09& 257 \\
 \object{ CS~29499-060  } & 13.03 & 11.82 & 6318 & 4.00 & 1.5 & -2.70 & 2.07 & 0.060 & 2.18 & 2.16 & 0.09& 163 \\
 \object{ CS~29506-007  } & 14.18 & 12.92 & 6273 & 4.00 & 1.7 & -2.91 & 2.05 & 0.040 & 2.15 & 2.15 & 0.09& 223 \\
 \object{ CS~29506-090  } & 14.33 & 13.06 & 6303 & 4.25 & 1.4 & -2.83 & 1.85 & 0.050 & 2.12 & 2.10 & 0.09& 222 \\
 \object{ CS~29518-020  } & 14.00 & 12.70 & 6242 & 4.50 & 1.7 & -2.77 & 2.10 & 0.110 & 2.14 & 2.13 & 0.09& 105 \\
 \object{ CS~29518-043  } & 14.57 & 13.37 & 6432 & 4.25 & 1.3 & -3.20 & 1.72 & 0.110 & 2.17 & 2.14 & 0.09& 107\\
 \object{ CS~29527-015  } & 14.25 & 13.02 & 6242 & 4.00 & 1.6 & -3.55 & 1.86 & 0.060 & 2.07 & 2.08 & 0.09& 121 \\ 
 \object{ CS~30301-024  } & 12.95 & 11.64 & 6334 & 4.00 & 1.6 & -2.75 & 1.77 & 0.060 & 2.12 & 2.10 & 0.09& 183 \\
 \object{ CS~30339-069  } & 14.75 & 13.47 & 6242 & 4.00 & 1.3 & -3.08 & 2.04 & 0.110 & 2.13 & 2.13 & 0.09& 117\\
 \object{ CS~31061-032  } & 13.90 & 12.59 & 6409 & 4.25 & 1.4 & -2.58 & 2.10 & 0.060 & 2.25 & 2.22 & 0.09& 116 \\
\hline
\hline
\end{tabular}
\end{table*}

\begin{table*}
\caption{Parametric fits in the A(Li)--[Fe/H] plane, 17 points}
\label{fits}
\begin{tabular}{ll}
\hline
\hline
&       Method    \\

\hline
\hline
$\rm A(Li) = 3.20(\pm 0.37)+0.37(\pm 0.13)\times[Fe/H] $& BCES\cr
$\rm A(Li)= 2.76 \pm 0.33)+0.22(\pm 0.11)\times[Fe/H] $&{\tt fitxy
\phantom{e}} $\chi^2 = 10$ ; P = 0.78\cr
$\rm A(Li) = 2.92(\pm 0.40)+0.28(\pm 0.13)\times[Fe/H]$ & {\tt fitexy}
~$\chi^2 = 9.5$ ; P = 0.85\cr
\hline
\hline
\end{tabular}
\end{table*}

\subsection{Dispersion in the plateau}

After removing the two stars discussed above, the sample of TO stars we can use
to probe the {\em Spite plateau} comprises 17 stars. The straight mean A(Li) of
the sample is A(Li) $= 2.11 \pm 0.094$ (s.d.). The error budget is totally
dominated by the error on \teff and totals about 0.09 dex, even if we adopt an 
error of 0.1pm for the EW of the Li doublet, thus leaving no room for any intrinsic scatter in the plateau. Of course, this relies on our estimate
of the error on $T_{\rm eff}$; 
if the true errors on \teff are smaller, a 
small amount of intrinsic scatter in A(Li) cannot be completely ruled out.

In their analysis of the {\em Spite plateau}, \citet{BM97} corrected the
measured Li abundances for the effects of both standard depletion and NLTE.
Similarly corrected values for our present sample are listed in Col. 9 of Table
\ref{tabli}. Both effects are rather small, due to the high effective
temperatures of our TO stars. With the corrections, the mean A(Li) lowers
slightly to A(Li) = 2.10, but there is hardly any change in standard deviation
(0.087 dex vs. 0.094 dex). Thus, the statistical properties of the sample change
very little whether we consider the measured A(Li) or the ``corrected'' A(Li). The following discussion refers throughout to the ``corrected'' A(Li), which 
we simply call A(Li).
It is suggestive that the dispersion is slightly reduced 
when the NLTE and depletion corrections are included, which suggests 
that those corrections are not far from the truth.

\subsection{The A(Li)--[Fe/H] plane\label{lifeh_sec}}

Figure \ref{fig_fehli} shows A(Li) versus [Fe/H] for our final sample of 17 stars.
At first glance, a slope of A(Li) with [Fe/H] seems to exist. However,
Kendall's $\tau$ test yields a probability of correlation between these
variables of 87\%; usually, correlations with a probability of less than 95\%
are not considered real. For example, if we remove \object{CS~31061-032} (the
most metal-rich star) from the sample and recalculate the statistic, the
correlation probability drops to 66\%.
 
Table \ref{fits} shows the results of several parametric fits to the data,
including the BCES algorithm \citep{ab}, a least-squares fit with errors in the
dependent variable only ({\tt fitxy}, \citealt{nr}), and a least-squares
algorithm with errors in both variables ({\tt fitexy}, \citealt{nr}). The BCES
and {\tt fitexy} fits formally indicate a possible slope, but only at slightly
more than 3$\sigma$ significance, confirming the negative result of the
non-parametric test discussed above. It is interesting to note that a simulation
of 10000 bootstrap samples, extracted from the real data set, fitted with BCES,
provides a mean slope of 0.439, 
but a standard deviation of 0.387, indicating that the slope is driven by a few
data points. When the most deviant points are excluded from the bootstrap
sample, essentially no slope is detected. 
A preliminary analysis of this sample was presented at the 2003 IAU General 
Assembly \citep{BIAU}, where we reported the detection of a similar slope 
to that obtained by {\tt fitexy}. 
The main difference 
in comparison to 
the former analysis is the removal from the sample of a few stars.

\subsection{The A(Li) -- \teff plane.}

\begin{figure}
\resizebox{\hsize}{!}{\includegraphics[clip=true]{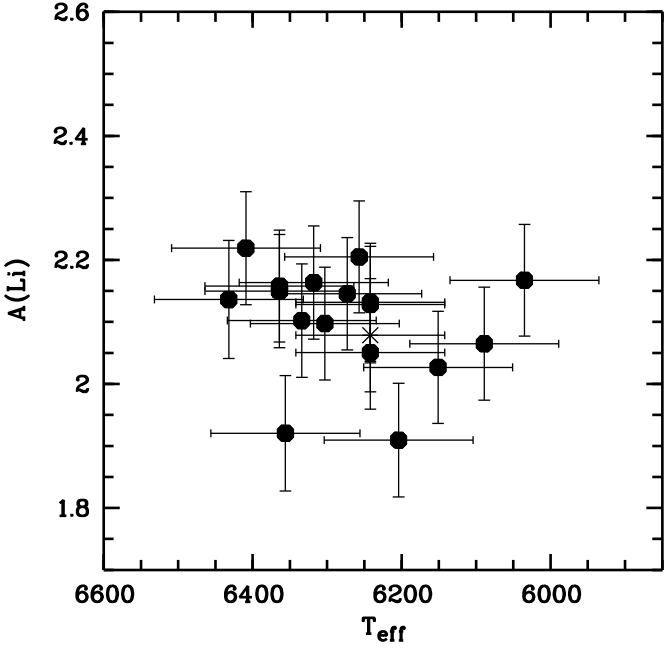}}
\caption{Derived Li abundances, corrected for standard depletion and
NLTE effects, versus $T_{\rm eff}$. The asterisk denotes the double-lined spectroscopic
binary \object{CS~29527-015}.
}
\label{fig_teffli}
\end{figure}

Figure \ref{fig_teffli} shows A(Li) versus \teff  for our sample. There is no
clearly detectable slope in this case. Kendall's $\tau$ provides a probability
of 87\% of a positive correlation. Neither of the two parametric fits detects a
slope at greater than 2$\sigma$ significance.

\begin{table*}
\caption{Parametric fits in the A(Li)--\teff  plane}
\label{fitsteff}
\begin{tabular}{ll}
\hline
\hline
&       Method    \\

\hline
$\rm A(Li) = -9.0(\pm 43)+0.18(\pm 0.69)\times (T_{eff}/100) $& BCES\cr
$\rm A(Li)=  1.07(\pm 1.32)+0.02(\pm 0.02)\times (T_{eff}/100) $
&{\tt fitxy\phantom{e}} $\chi^2 = 14$ ; P $= 0.53$\cr
$\rm A(Li) = -1(\pm 3)+0.05(\pm 0.04)\times(T_{eff}/100)$ & {\tt fitexy}
$\chi^2 = 13$ ; P = 0.62\cr

\hline
\hline
\end{tabular}
\end{table*}

\begin{figure}
\resizebox{0.9\hsize}{!}{\includegraphics[clip=true]{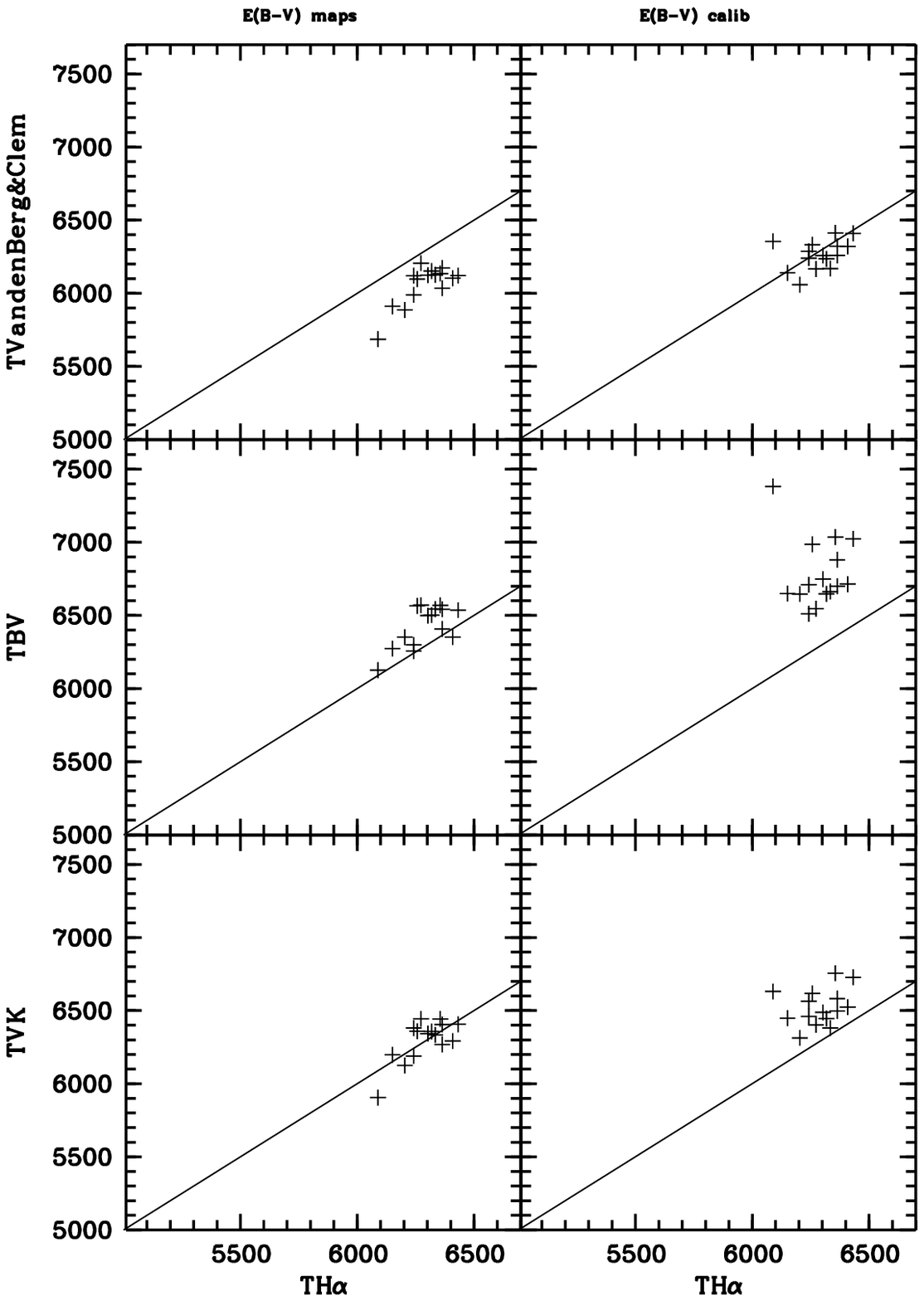}}
\caption{Comparison of different colour temperatures with the 
H$\alpha$ scale. From top to bottom we show the temperatures
derived from $V-K$, and the \citet{alonso} 
calibration, $B-V$ using the \citet{clem} theoretical colours,
and $B-V$ and the \citet{alonso} calibration.
The panels on the left adopt the reddening derived
from the \citet{schlegel} maps, corrected as in
\citet{BMB00}. The panels on the right adopt
the reddenings derived from the \citet{BCM} intrinsic colour calibration. In
each panel the one-to-one relation is shown as a solid line. }
\label{tscales}
\end{figure}

\begin{figure}
\resizebox{0.9\hsize}{!}{\includegraphics[clip=true]{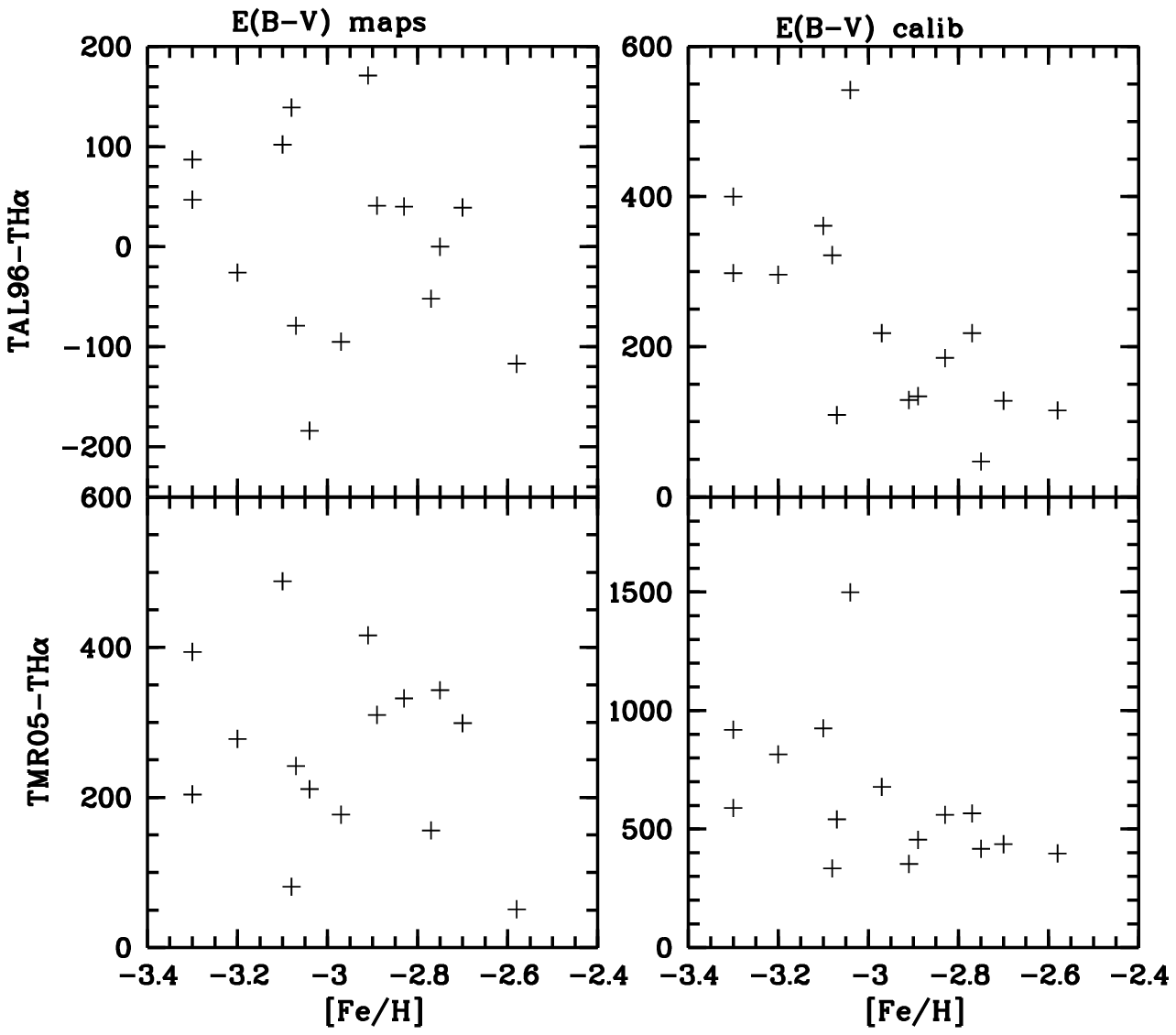}}
\caption{Comparison of the different effective temperature
scales based on $V-K$. The lower panels show the difference
from the H$\alpha$ based temperature of the
\citet{RM05} calibration. The upper panels show the same 
for the \citet{alonso} calibration. Note the different vertical
scales in each panel. The panels on the left assume the reddening
derived from the \citet{schlegel} maps, while those on the
right use the reddening that was 
deduced from the \citet{BCM} calibration.}
\label{vkscales}
\end{figure}

\subsection{\label{TEFF} Effects of different \teff scales}

The slope of A(Li) versus [Fe/H] described in Sect. \ref{lifeh_sec}, 
detected at slightly over
3$\sigma$, is of the order of 0.4 dex/dex, more than a factor of two larger than
that reported by \citet{ryan}. A lively debate exists, both about the reality of
this slope and, if real, about its interpretation. \citet{B2002} re-analysed a
subsample of the stars from \citet{ryan} for which he had accurate IRFM
temperatures, but was unable to detect any slope using the published Li
equivalent widths and metallicities. Recently, \citet{melendez} analysed a sample
of 62 metal-poor dwarfs from the literature, using IRFM effective temperatures,
and again found no detectable slope. This result might be taken as confirmation
of the conclusions of \citet{BM97}. Note that the offset in mean A(Li) of the
{\em plateau} between \citet{BM97} and \citet{melendez} is due essentially to
the different model atmospheres employed (ATLAS overshooting models by
\citealt{melendez} vs. ATLAS non-overshooting models by \citealt{BM97}). Thus, a
slope may appear or disappear, depending on the temperature scale used. In this
paper we have used H$\alpha$-based effective temperatures, which are usually
found to be on the same scale as IRFM temperatures \citep{gratton,barklem02}.

The effect of the adopted temperature scale on the existence of a slope (and the
mean value of A(Li) on the {\em plateau}) requires closer inspection. For most
of our stars we had $JHK$ magnitudes from 2MASS\footnote{\href{http:
//pegasus.phast.umass.edu/}{http://pegasus.phast.umass.edu/}} as well as $UBV$;
thus, we may consider four temperature sensitive colours: $J-H$, $J-K$, $B-V$, and
$V-K$. When deducing temperatures from colours one is always confronted with the
problem of reddening. We decided to explore two different approaches, 
to be able to estimate the uncertainty on the reddening also. We used the
$E(B-V)$ derived from the reddening maps of \citet{schlegel}, corrected as in
\citet{BMB00}, as well as the $E(B-V)$ derived from the \citet{BCM} intrinsic
colour calibration. The latter requires KP and HP indexes, which were available
for all of our stars from the medium-resolution HK-survey spectra. 

We note that the \citet{BCM} relation is derived for stars that are more
metal-rich than [Fe/H] = --2.5. Thus, for all our stars we are applying an
extrapolation beyond its stated range of validity. It is nevertheless
interesting to compare these two $E(B-V)$ values for each star. On average, the
\citet{schlegel} maps yield a reddening 0.05 mag smaller than the
alternative calibration (the dispersion about the mean is of similar size as the
offset, $\sim $ 0.04 mag).

\citet{BCM} noted that an offset of 0.01 mag exists between their calibration
and the reddening derived from the \citet{schlegel} maps. In this case the
offset appears considerably larger; however, the dispersion is compatible with
the expected accuracy of each method, which is $\sim$0.02 mags for the maps and
0.03 mags for the calibration.
Using these two values for the reddening, we derive effective temperatures from
the \citet{alonso} calibrations for all four colours. The \citet{alonso}
calibration for $V-K$ is given in the Johnson system; we therefore used the
transformation given in \citet{cutri} to transform the 2MASS $K$ magnitude to the
\citet{BesselBrett} homogenized system. The $J-H$ and $J-K$ calibrations are
given for the TCS system; since a direct transformation 2MASS to TCS is not
available, we performed a two-step calibration: 2MASS to CIT using the
transformation of \citet{cutri}, and then CIT to TCS using the transformation of
\citet{Alonso94}. For $B-V$ we determined \teff using both the \citet{alonso}
calibration and the theoretical colours of \citet{clem}. 

The IRFM \teff 
scale has been recently revised by \citet{RM05}, who added a few
metal-poor stars to the original sample of calibrators of \citet{alonso}, and
computed new polynomial fits. We also considered the $V-K$ calibration of
\citet{RM05}; in this case the calibration is performed assuming $V$ in the
Johnson system and $K$ in the 2MASS system. Figure \ref{tscales} shows a comparison 
of some of the colour-based effective temperatures with those derived from
H$\alpha$. The plot suggests that offsets exist among the temperatures derived
from different colours, for any chosen $E(B-V)$, as well as with respect to the
$H_\alpha$ temperature. For some choices the colour-based
\teff and the H$\alpha$ temperature appear in good agreement; this is the case
for both the $V-K$ temperature with the 
\citet{schlegel} reddenings and the 
\teff obtained from $B-V$ with the \citet{clem} colours and the \citet{BCM} 
reddening corrections.   

We wish to look more closely at the {\it V-K} calibrations of \citet{alonso} and
\citet{RM05}. For this purpose, from the sample of 17 stars we exclude 
BS~16968-061, which has no 2MASS photometry, and BS~17570-063, 
which  lacks some of the spectral 
information needed by the \citet{BCM} calibration.
With the \citet{BCM} reddenings, we note that
both the \citet{alonso} and the \citet{RM05} calibrations provide higher
temperatures than those estimated from H$\alpha$, \citet{RM05} being by far the
hottest. Furthermore, the differences are larger for the more metal-poor stars,
as clearly seen in Fig. \ref{vkscales}. 

On the other hand, when the
reddening based on the \citet{schlegel} maps is adopted, there appears
to be no trend with metallicity. In this case the mean difference 
$\rm T_{{(V-K)}_{A96}}-T_{H\alpha}$ is only 7.5~K with a standard deviation of 100~K, as compared to a mean difference 
$\rm T_{{(V-K)}_{RM05}}-T_{H\alpha}$ of 265~K, with a standard
deviation of 122~K. None of the above discussed residuals shows any trend with
\teff. We conclude that the IRFM-based temperatures derived from the
\citet{alonso} calibration are in good agreement with the H$\alpha$ temperatures,
even for these extremely low metallicities, in keeping with what is found at
higher metallicities \citep{gratton,barklem02}. On the other hand, the
temperatures derived from the \citet{RM05} calibration are considerably higher
and essentially incompatible with the H$\alpha$ temperatures. These 
discrepancies suggest that a systematic error in the adopted temperature 
scale of the order of 200~K is still possible. 

\begin{figure}
\resizebox{\hsize}{!}{\includegraphics[clip=true]{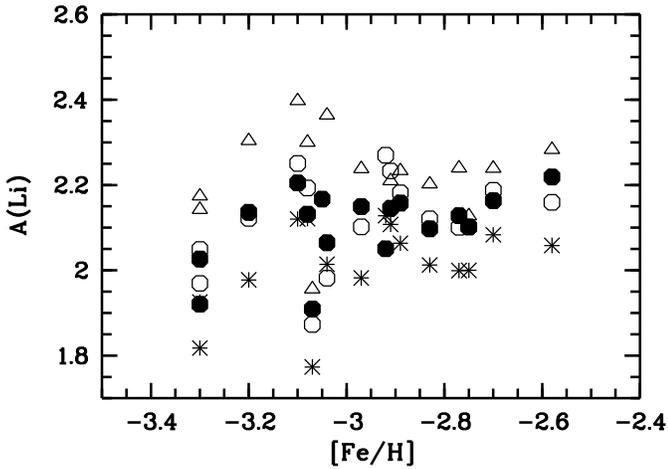}}
\caption{
Effect of different temperature scales on the {\em Spite plateau}. 
Filled circles: H$\alpha$ scale; open circles: $V-K$ with reddening 
from the \citet{schlegel} maps; open triangles: 
$V-K$ with reddening from the \citet{BCM} calibration;
asterisks: $B-V$ from \citet{clem} and reddening from the \citet{schlegel}
maps. }
\label{tplateau}
\end{figure}

To illustrate the effects of different temperature scales on the {\em Spite
plateau}, Fig. \ref{tplateau} compares A(Li) as derived with the H$\alpha$
temperatures with that derived using three other \teff scales. The comparison is
slightly inconsistent, since we did not recompute [Fe/H] with the different
\teff scales, only A(Li). However, we believe this is sufficient to illustrate
the general trends. Figure \ref{tplateau} shows that, although the $V-K$
temperatures (with the \citet{schlegel} reddening) and the 
H$\alpha$ temperature (which should not be
affected by reddening) appear to be on the same scale, the {\em Spite plateau}
has a very different appearance in the two cases. With the H$\alpha$-based
temperatures there is (weak) evidence for a slope in A(Li) vs. [Fe/H]; when
using the $V-K$ temperatures there is no evidence for a slope (but the scatter in
A(Li) is larger in this case). The most extreme situation is when we adopt the
\citet{RM05} $V-K$ calibration and the reddenings from the \citet{BCM}
calibration -- not only is there no slope (probability of correlation of 2\%),
but two stars have A(Li) $\sim 2.6$, and one star would be assigned essentially
a meteoritic Li abundance. This is a consequence of the fact that in this case
the temperature difference with respect to the H$\alpha$ scale is larger at
lower metallicities. This trend may be totally spurious, and arise from the fact
that we are extrapolating the \citet{BCM} calibration beyond its range of
validity.

We note here that our H$\alpha$ temperatures yield an excitation equilibrium for 
iron, i.e., no detectable abundance trend with excitation potential, for most
of our stars (13 out of 19). For those in which a mild slope was found (at most
0.06~dex/eV), it could be removed by adopting a slightly {\em cooler} \teffns, 
by 100~K in the worst cases. In contrast, with the high \teff derived from 
the \citet{RM05} calibration, no iron-excitation equilibrium is achieved;
remaining slopes are of the order of 0.15~dex/eV. Furthermore, achieving 
iron-ionization equilibrium in such cases would require substantially
larger surface gravities, as large as log~g = 5.5 in some cases, which is 
very unlikely in TO stars.
We conclude that the \citet{alonso} calibration is in good agreement with
both H$\alpha$ temperatures and iron excitation temperatures.

\subsection{Comparison with the results of Asplund et al.}

In a recent paper \citet[][, hereafter A06]{asplund} 
measured both $^6$Li and $^7$Li from UVES
spectra for a sample of 24 metal-poor stars. To investigate $^6$Li,
which has a very small isotopic separation from $^7$Li, the S/N ratios of their
spectra had to be extremely high, and therefore their measured equivalent widths
are of very high accuracy as well. Their adopted temperature scale is based on
the wings of H$\alpha$, modelled with the \citet{barklem} broadening theory, and
is thus, in principle, identical to ours. 

Since we have no stars in common with A06, and at the
suggestion of the referee, we downloaded UVES spectra from the ESO archive for
two of the most metal-poor stars in the sample of A06, LP~815-43 and
CD --33 1173. These data include the spectra used by A06, taken with
the image slicer, as well as other spectra taken with a slit, which are more
similar to our own data. The details of the analysis of H$\alpha$ from these
data is shown in the appendix, and the spectra used are detailed in Table
\ref{archivedata}. The main result is that, even taking into account the
different data available and the differences in data reduction and line-profile
fitting, our temperature scale agrees with that of A06, although a
zero point shift of up to to 40 K is possible. To be certain on this point,
we should determine effective temperatures for a large fraction, if not all of, the stars in
the A06 sample. This is clearly beyond the scope of the present
paper. 

We also  used these data to compare the metallicity scales.  Table
\ref{fe_LP815_43} lists our measurements of iron lines for LP~815-43. These
data, analysed with the \teff = 6400~K used by A06 and our models and atomic
data, provides [Fe/H]=--2.94, log~g = 3.90, and $\xi=1.6 $ \kms. We conclude that
the metallicity scales of the present paper (based on neutral iron lines) and that of
A06 (based on ionized iron lines) are offset by $\sim 0.2 $dex, in the
sense that our analysis provides lower metallicities. 
A06 find a mean difference between A(\ion{Fe}{i}) and 
A(\ion{Fe}{ii}) of 0.08 dex, which suggests that the offset 
between the metallicity scales is not entirely due to the
use of neutral or ionized iron, but may also be related to 
differences in the choice of lines, the adopted atomic data, 
and, possibly, also the line formation codes.
It is interesting to note, however, that the study 
of metal-poor subgiants by \citet{ana}, who adopt
the same approach and methods as A06,
finds a mean imbalance of 0.19 dex between neutral
and ionized iron, virtually identical to the 
offset we are discussing.
It may well be that, after all, the line-to-line
scatter for both \ion{Fe}{i} and \ion{Fe}{ii}
is too large to allow strong conclusions 
on the ionization balance, and, consequently, on surface gravities.
Whatever the case, for  the purpose of the
present comparison we applied this shift to the A06. Although a
thorough re-analysis of their spectra would be preferable, again, it would be
beyond the purpose of the present paper.

\begin{figure}
\resizebox{\hsize}{!}{\includegraphics[clip=true]{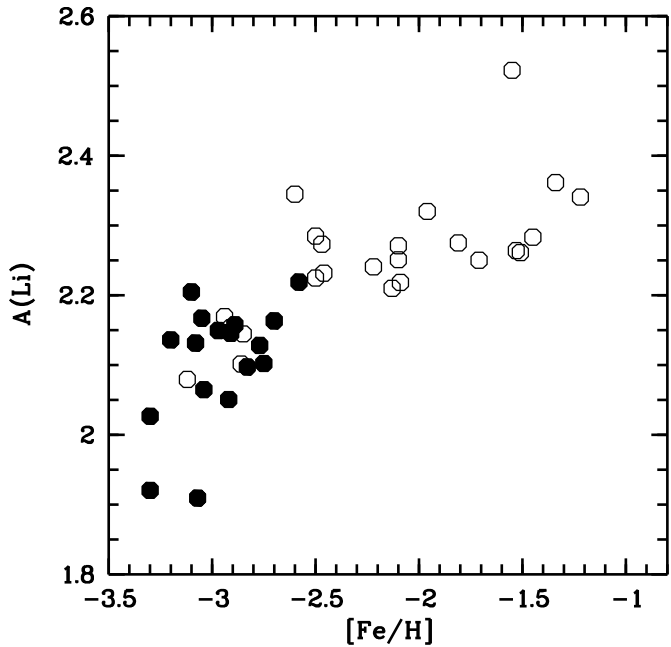}}
\caption{
Comparison of our sample (filled circles), with
that of A06 (open circles).
The temperature scale of both samples is based on H$\alpha$ profiles.
}
\label{asplund}
\end{figure}

We finally compared the derived A(Li) abundances. Somewhat to our surprise, we
found that using the equivalent widths published by A06, and their
adopted atmospheric parameters, with the use of {\tt turbospectrum} the derived
lithium abundance is about 0.04 dex higher than the one found by the A06
study. In spite of extensive investigations, with the collaboration of M.
Asplund, we were unable to resolve the reason of this discrepancy. We did,
however, ascertain that this small offset is not due to the models employed, 
to the adopted atomic line data, or to the adopted continuum opacities. This
finding indicates the likely level of systematic error in derived Li abundances
due to the use of different spectrum synthesis codes. For the present comparison
we recomputed all of the Li abundances using the equivalent widths and model
parameters of the A06 study.
The A06 ``rescaled'' data is plotted in Fig.\ref{asplund}, together
with our own. The fact that now the most metal--poor stars of the A06
sample fall in the midst of our measurements supports the notion that the adopted
rescaling in [Fe/H] and the re-computation of A(Li) have brought the two samples
onto the same scales.

\begin{figure}
\resizebox{\hsize}{!}{\includegraphics[clip=true]{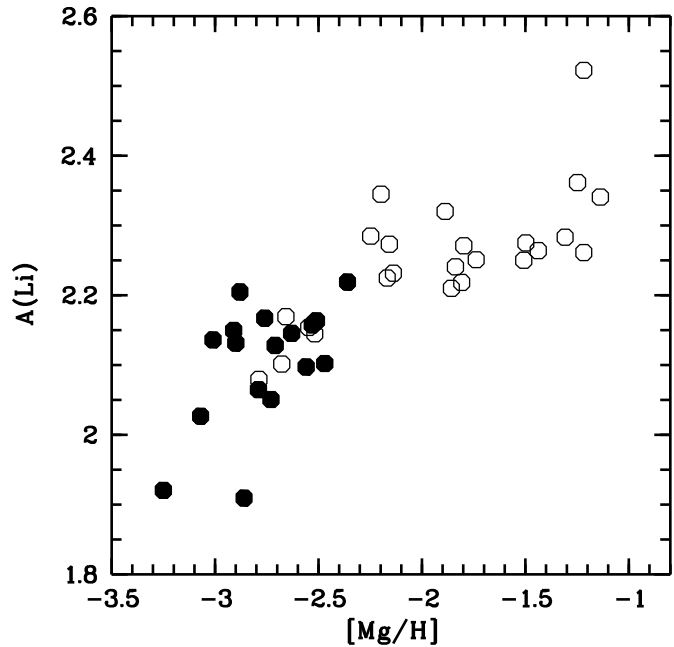}}
\caption{
Comparison of our sample (filled circles), with
that of A06 (open circles) using Mg as
a reference element.
For the stars of A06 Mg has been estimated
by rescaling their oxygen measurements (see text).
}
\label{asplundmg}
\end{figure}

As was pointed out by \citet{cayrel04}, 
Mg is perhaps a better reference element
to study chemical evolution, since 
its production in relatively external layers of massive stars
is linked to the volume of the Li-poor external layers more
directly than to the volume of the deep iron-rich
ejected layers, which depend on additional parameters (fallback and
mass cut).
In Fig. \ref{asplundmg} we plot A(Li) as a function
of [Mg/H]. Magnesium has typically been measured using 
7 lines for each star, and should
be quite accurate; details will be given in a 
forthcoming paper of this series
(Spite et al., in preparation, see also \citealt{iaus05}). 
For the A06 data, we estimate Mg abundances from their 
[O/H] measurements by adopting 
[Mg/H] = [O/H]--0.46, since we find a mean [Mg/O]= --0.46 
for our giants \citep{cayrel04}, and [Mg/O] is constant with [Mg/H] 
(see Fig. 13 of Cayrel et al. 2004).

\begin{figure}
\resizebox{\hsize}{!}{\includegraphics[clip=true]{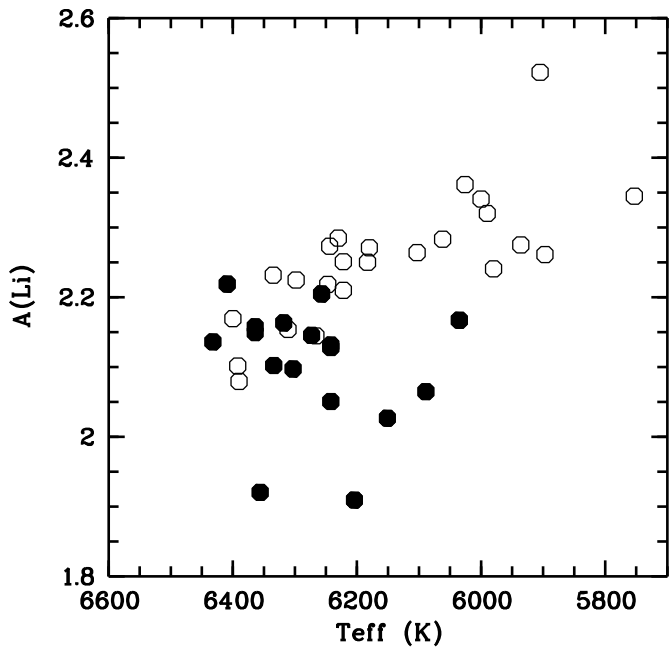}}
\caption{
Comparison of our sample (filled circles), with
that of A06 (open circles) in the Li-\teff plane.
}
\label{asplundteff}
\end{figure}

Figure \ref{asplundteff} displays A(Li) versus \teff for the two samples. It is
clear that, while our data do not exhibit any slope with \teffns, the data of
A06 exhibit a sizeable slope of about $-0.04$ dex/100~K, the
hotter stars exhibit the lowest Li abundances. This may be related to the 
structure of the A06 stellar sample, 
which shows a strong correlation 
between [Fe/H] and \teffns,   their most metal-poor
stars  are also the hottest. We stress that the A(Li) vs. \teff 
slope is in
the {\em opposite} direction of that found by \citet{ryan96}. Note as well
that for the most
metal-poor stars of the sample described by \citet{CP05}, there exists a slope similar
to that of Asplund et al., although it is not very well defined. Also, \citet{ryan01}
provide a figure of A(Li) versus \teffns, which suggests a 
changing slope for the hottest stars, but only as a subtle effect.

\section{Discussion}

The two main reasons that drove our investigation were, first, to understand
whether the interpretation of a primordial origin of the {\em Spite plateau} is
supported by the new observations of the most metal--poor stars, and secondly,
if this is the case, to assess the value of the primordial Li abundance. In
this context, the existence or non-existence of a slope in A(Li) versus [Fe/H]
in the {\em Spite plateau} plays an important role. It is possible that a real
slope may indicate either that there has been Li production by cosmic rays in
the early Galaxy, as suggested by \citet{ryan00}, or that some atmospheric
phenomenon, such as diffusion, has altered the atmospheric Li abundance in a
metallicity-dependent manner, or even that a metallicity-dependent
astration of Li in the progenitors, as proposed recently by \citet{piau}, may
exist. In the first case the primordial Li abundance might be estimated by
simply extrapolating the slope down to very low metallicities, while in the
other cases the primordial abundance cannot be estimated in a model-independent
way.

\subsection{The scatter in the plateau}

As has been found by all (recent) investigations, based on data of the highest
quality, we confirm a very low scatter
in the Li abundances among stars on the Li plateau, a scatter which may be
explained by observational errors alone. However if we arbitrarily divide the
sample of 17 stars into two subsamples, one with [Fe/H]$\le-3.0$ (8 stars) and
its complement (9 stars), we find a scatter of 0.11 dex for the lower-metallicity
subsample and 0.05 dex for the higher-metallicity subsample. The increased scatter in
A(Li) for the lowest metallicities could be due to the fact that they are TO
stars. The transition between the dwarf and subgiant phase may produce some
transport processes that result in a reduction of photospheric Li in these
stars, but this remains to be confirmed by further study, and empirical
data suggest the contrary \citep{CP05}.

\subsection{The slope of the plateau}

From our measurements alone the evidence for any slope in A(Li) vs. [Fe/H] in
the {\em Spite plateau} is weak. Its existence or non-existence remains a very
delicate issue, the resolution of which will likely require temperatures with
accuracies of the order of 50~K, roughly a factor of two better than what can be
achieved at present. The larger scatter in A(Li) when the $V-K$ temperatures are
adopted can be ascribed to a larger error on estimated \teffns, as compared to
that obtained using the H$\alpha$-based temperatures. We attribute this error to  
being dominated by the uncertainty in the reddening, which will always prevent
one from obtaining accurate temperatures from colours alone. 

The situation is totally different when we look at the combined sample 
formed from our stars and those of
A06. In this case, sizeable slopes exist, both with
\teff and with [Fe/H], which are, however, entirely driven by the A06
data at higher metallicity. We note that while A06. stress the
existence of a slope with [Fe/H] in their data, they do not comment or even
mention the existence of a steep slope with \teffns. 

We are thus faced with surprising
and somewhat contradictory information. The correlation between [Fe/H] and
A(Li) is clearly present, with a slope of about 0.15 dex/dex. The impression
one obtains from inspection of Fig. 7 is that of an almost vertical drop of
lithium abundance at the lowest metallicities. The fact that the
A06 data also displays a steep slope with \teffns, in the opposite
direction of that which has been found by previous investigations, suggests
that A06 may have unveiled a new physical 
phenomenon for the first time 
(see Sects. 6.5 and 6.7 for possible interpretations). This is, of
course, predicated on the absence of a systematic temperature-dependent error in
the adopted \teff scale. 

Note that for LP~815-43, which is one of the stars
that outline the slope, \citet{RM06} derive \teff = 6622~K, 222~K higher than
the $H\alpha$-based \teffns. If one were to adopt this \teffns, the derived A(Li)
would rise to 2.3, i.e., at the level of the less metal--poor stars. It is clear
from figure 8 of \citet{RM06}, and confirmed by our own analysis, that their
temperatures are considerably hotter than H$\alpha$-based temperatures only for
the extremely metal-poor stars. Our analysis disfavours the \citet{RM05}
temperature scale, on account of its inconsistency with the iron excitation
equilibrium. However, it is unclear at this stage whether the increase in the
difference between the \citet{RM05} \teff estimates and H$\alpha$-based \teff
estimates with decreasing metallicity reflects a problem in the H$\alpha$ scale,
in the \citet{RM05} scale, or both.

\begin{figure}
\resizebox{\hsize}{!}{\includegraphics[clip=true]{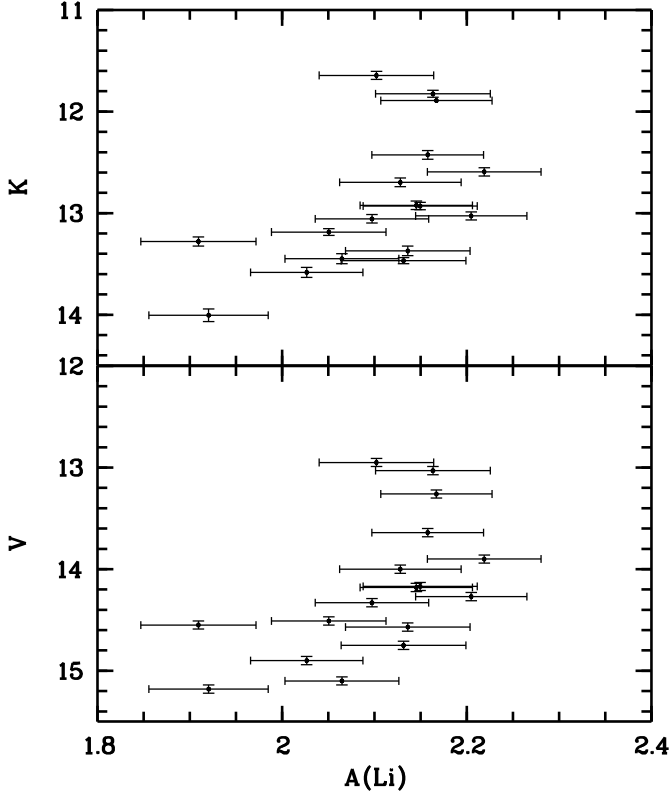}}
\caption{Li abundance vs. apparent $V$ and $K$ magnitudes. Kendall's 
$\tau$ test indicates a correlation at over the 99\% level in both cases.
}
\label{fig_Vli}
\end{figure}

\begin{figure}
\resizebox{\hsize}{!}{\includegraphics[clip=true]{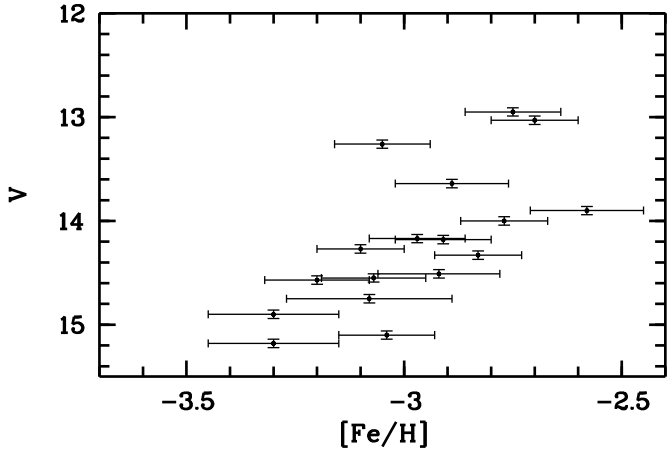}}
\caption{[Fe/H] vs. apparent $V$ magnitude. 
Kendall's $\tau$ test indicates a correlation at over the 99\% level.
\label{fig_Vfeh}
}
\end{figure}

We find it somewhat disturbing that, at least in our present sample of stars,
a slope of A(Li) with {\em apparent magnitude} exists, as shown in Fig.
\ref{fig_Vli}. The slope is obvious, whether one employs $V$ or $K$ magnitudes; the
fainter stars have lower derived Li abundances. The correlation is detected at
over the 99\% confidence level for both bands, and hence cannot be questioned. A
correlation, with a similar confidence level, exists between apparent magnitude
and [Fe/H] (Fig. \ref{fig_Vfeh}). 
It is worth mentioning that this correlation between apparent 
magnitude and A(Li) is not peculiar to our sample. The 62 
stars in 
\citet{CP05} with \teff$> 5700$~K and [Fe/H] $\le -1.5$ exhibit a similar trend.

The correlation between [Fe/H] and apparent magnitude might be understood as an
observational bias, such as that suggested by \citet{B2002}. To observe
a similar number of stars at [Fe/H] = --3.2 as at [Fe/H] = --2.8, one must
sample a larger volume of the Galactic halo. As a result, on average, the most
metal-poor stars are more distant, and their apparent magnitudes are fainter.
One can easily see that if such a correlation exists {\em along} with
a correlation between A(Li) and [Fe/H], {\em then} a correlation between A(Li) and
apparent magnitude must exist. Moreover, the slopes in the different planes must
be simply related, e.g., the slope in the [Fe/H], A(Li) plane must be the
product of the slopes in the $K$, A(Li) and $K$, [Fe/H] planes, which is in fact
verified.

\citet{B2002} also suggested that the trend of A(Li) vs. [Fe/H] may be due to 
observational bias. In a large sample of stars from the literature, he found a
clear trend of the equivalent width of the Li doublet with metallicity. This
arises because the most metal-poor stars are also the hottest. The combination
of small equivalent widths and faint magnitudes may yield less accurate
measurement and systematic underestimates of the equivalent widths.  

In \citet{B2002}, the slope of A(Li) vs. [Fe/H] was seen in the full sample of
73 stars, but not in the subsample of 22 stars with reported errors in
equivalent widths less than 0.1 pm. The present sample has been designed to
avoid such a bias, both because all of the stars are rather hot and because we
obtained very high S/N ratio for all stars, independent of their apparent
magnitude. Our sample does {\em not} exhibit any trend of the Li doublet
equivalent width with metallicity.  In our sample, 15 of 17 stars have errors on
the equivalent width of the Li doublet less than 0.1 pm; the two remaining stars
have errors of 0.11 pm, which shows that our goal of obtaining a uniformly high
S/N ratio for all our program stars has been achieved.

The situation is therefore the following. Having established that there exists a
plausible reason for the existence of a correlation between [Fe/H] and apparent
magnitude (the observational bias), at least one of the two correlations A(Li)
vs. apparent magnitude and A(Li) vs. [Fe/H] must have a physical origin. The
other may simply be a consequence of the former two.

\subsection{Li production by Galactic cosmic rays}

If the trend with [Fe/H] is real, we must try to understand the physical
reason for it. \citet{ryan00} interpreted the correlation found by them as
evidence for Galactic production of Li. However, adopting this
interpretation results in a primordial Li abundance in stark contradiction 
to SBBN (with $\eta$ derived from the WMAP observations). Extrapolating the
trend of A(Li) linearly down to a metallicity of [Fe/H] = --4.0, we find A(Li)
=1.80, i.e., (Li/H) $=6.3\times10^{-11}$, compared to the minimum (Li/H) $\sim
1.13\times10^{-10}$ predicted by our Kawano-code primordial nucleosynthesis
computations. To avoid finding a primordial Li {\em below} the minimum allowed
by the theoretical computations, \citet{ryan00} increased all their A(Li) values
by 0.08 dex, arguing that this brought the zero point of their temperature scale
in agreement with the IRFM temperatures.

\begin{figure}
\resizebox{\hsize}{!}{\includegraphics[clip=true]{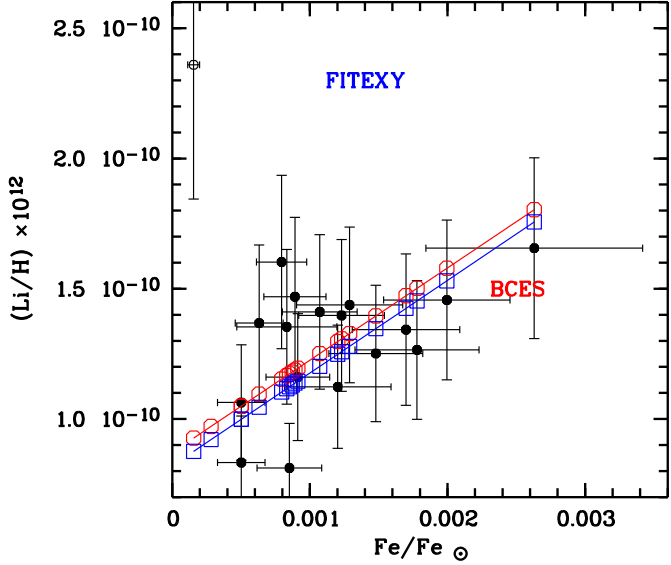}}
\caption{Fits of (Li/H) as a function of $\rm (Fe/Fe_\odot)$.
}
\label{fitlin}
\end{figure}

\citet{ryan00} argued that an early flux of Galactic cosmic rays (GCR) should 
result  in a lithium  
production that exhibits a linear trend of (Li/H) with $\mathrm{
(Fe/Fe_\odot)}$. In this case a fitting function of the form $\mathrm{ (Li/H)
=a+b(Fe/Fe_\odot) }$ is more appropriate than a linear fit of the logarithmic
abundances, as was discussed above. 
Figure \ref{fitlin} shows the result of such a fit. The BCES fit and {\em fitexy}
provide similar results. BCES suggests  $\mathrm a= (8.72\pm1.71)
\times10^{-11}$ and $\mathrm b=(3.54\pm1.10)\times 10^{-8}$. 
Extrapolation down to a metallicity of [Fe/H] = --4.0 implies (Li/H)
$=8.72\times10^{-11}$, which is higher than the logarithmic fit, but still below
the minimum predicted by SBBN ((Li/H) = $1.13\times10^{-10}$). It is clear that
for all metallicities below [Fe/H] $= -4.0$, the term $\mathrm{ b(Fe/Fe_\odot)}$ is
negligible compared to $\mathrm a$, and if we take the $\mathrm a$ value at
$+3\sigma$ we have (Li/H)$\sim 1.61\times10^{-10}$. 
We cannot avoid the conclusion that if the trend of A(Li) with [Fe/H] is due to
GCR Li production, then either SBBN is wrong and should be replaced by some
other theory of Li formation, or the baryonic density determined by WMAP is
overestimated and should be reduced to a lower value, compatible with both the
D/H and the Li/H values. 

\subsection{Diffusion at work?}

One alternative explanation would be that the correlations between A(Li) and
[Fe/H] could be caused by some metallicity- and/or temperature-dependent Li 
depletion
mechanism. In this regard, it is of significance that recent investigations of
the effects of diffusion (with the inclusion of turbulence: \citealt{ric1, ric3,
michaud,richard}) have been able to reproduce the Li-\teff dependence found by
\citet{ryan}. \citet{richard}, in their figure 3, show that their best-fitting
model provides a curve with a mean slope vs. \teff of 0.013 dex/100K. It is
interesting to recall here, that, while our sample alone shows no trend
with \teff, when including the sample of A06 we find a slope that is
considerably larger, but of the opposite sign. Could it be that we
are seeing the downturn of the {\em Spite plateau} at the highest temperatures,
predicted by the old diffusive models \citep[e.g.,][]{VC95}? Here the downturn could
be linked to metallicity rather than to temperature. 

No curve is provided by \citet{richard} to illustrate the ``constancy'' of
A(Li) versus [Fe/H], but the text of the paper emphasises the basic facts that
should lead to this constancy. The primary theme of \citet{richard} is not the
numerical values of the (small) slopes of A(Li) vs. \teff and [Fe/H], but the
global constancy of the modelled A(Li) on the plateau and the net depletion.
These authors claim to be able, by this approach, to satisfactorily close the
gap between the mean observed value of A(Li) of the plateau and the A(Li) value
derived from the WMAP measurements. 

Recently, \citet{Korn06} have claimed the detection of a diffusion ``signature''
in the Globular Cluster NGC 6397. According to their analysis the TO stars
in this cluster exhibit lower iron and lithium abundances than the slightly more
evolved stars. Both these facts may be interpreted in the framework of the
diffusive models of \citet{richard}. This result is very suggestive; however, it
relies heavily on the adopted temperature scale, which is plausible, albeit
inconsistent with the cluster photometry. An increase by only 100 K of the
effective temperature assigned to the TO stars would remove the abundance
differences between TO and subgiant stars. Previous analyses of the same
cluster \citep{castilho,gratton}, with different assumptions on the effective
temperature scale, failed to find any difference in [Fe/H] between TO,
subgiant, and giant stars. Thus, we are again faced with the need to be able to
determine effective temperatures to better than 50 K to confirm or
refute this result. 

\subsection{ A Population II Li dip?}

Another possibility is that we are observing a phenomenon that is similar to
the ``Li dip'' \citep{BT}, a depletion of lithium observed within a small 
temperature range in some field Pop I stars and also in old open clusters.
We recall that the Li dip is not observed in young clusters, but 
is seen in old, preferably slightly metal-poor clusters. 
Interpretations by diffusion \citep{michaud86,proffitt1990}, rotationally
induced mixing \citep{erika,talon,pin92}, or mass loss \citep{SSD} have been
proposed. 

The observation of a low Li content in subgiants lead \citet{Balachandran} 
to conclude that the Li dip is due to the previous destruction of Li in
the main sequence phase, which is 
improbable for very low-metallicity dwarfs. \citet{Balachandran} found 
that a lithium
depletion is noted in subgiants which had a temperature larger than 7000 K on
the main sequence, and suggested that this depletion corresponds to the dip. At
solar metallicity the dip is around the mass 1.35 $M_{\sun}$, at intermediate
metallicity it is around 1.25 $M_{\sun}$, and at low metallicity it is around 
1.1 $M_{\sun}$. A dip at 0.8 $M_{\sun}$ for extremely low metallicity would be 
in line with this trend. 

\citet{dearborn} have claimed that the mass-loss scenario \citep{SSD} that 
is capable of explaining the Hyades Li dip {\em necessarily} predicts an Li
dip for Pop II stars also. In their computation, the red edge of the dip should
appear at around \teff = 6600 K, i.e., only slightly hotter than our most
metal-poor stars. No Li dip has been clearly observed in Pop II stars yet;
however, Fig. \ref{asplundteff} is reminiscent of what one should see approaching
the Li dip from the cool side. So, perhaps, the temperature trend discovered by
A06 constitutes the first evidence in favour of a Pop II Li dip. It
is unclear why previous investigations have failed to detect it, as it is
unclear why it is not detected among Pop I field stars \citep{lambertreddy}.
Such a process would not necessarily explain the low Li in the 
hyper iron-poor
dwarf HE~1327-2326 \citep{frebel}, which does not have a very low global
metallicity $Z$ \citep{Collet}, but has a still lower mass.

\balance

\subsection{The $^6$Li plateau}

The measurements of $^6$Li in halo stars also have implications for the
interpretation of the {\em Spite plateau}. In the 1990s a firm detection of
$^6$Li was achieved for only one halo star, HD~84937 \citep{smith93,HT1994,
smith98,cayrel99}. The measured $^6$Li/$^7$Li ratio was of the order of 5\%.
Detections of $^6$Li have been claimed for only three other halo stars, but
these have been withdrawn on the basis of higher quality spectra.
\citet{smith98} claimed a
detection of $^6$Li in BD $+26^\circ 3578$ (= HD~338529) at the level of 5\%,
which has been changed to an upper limit by A06.  The detection of
$^6$Li in HD~140283 by \citet{DR2000} was changed to an upper limit by
\citet{aoki}.  Finally, the detection of $^6$Li in G~271-162 by \citet{nissen} was
changed to an upper limit by A06.

Recently, A06 claimed detection of $^6$Li in nine more halo stars.
Surprisingly, these stars seem to have the same Li isotopic ratio of $\sim $
5\%, defining a $^6$Li plateau which appears to mirror the $^7$Li plateau.
 
Given the extreme difficulty of measuring the $^6$Li ratio and the history of
several claimed detections and withdrawals, caution is advised in accepting the
measurements of A06 until they are confirmed by an independent
analysis, preferably based on data from a different spectrograph. However, the
fact that the $^6$Li measurement in HD 84937 has been confirmed in four
independent analyses suggests that it is unlikely that all the detections
claimed by A06 will later be found to be upper limits. Hence, for the
sake of the discussion, we shall take these measurements at face value.

The measurements of $^6\mathrm{Li}$ in halo stars imply the production of both
Li isotopes by GCR at [Fe/H]$\sim -2.3$. At low metallicities, the
$\alpha-\alpha$ fusion process may be competitive with spallation. The fusion
process is favoured in the special environments near supernovae; thus, unless
the early Galaxy was very well mixed, there should be some scatter in Li, 
and some
stars having higher Li because they were formed from an ISM enriched by
$\alpha-\alpha$ fusion. This could explain a tiny scatter; moreover, since on
average it is more likely to observe such a process at higher metallicities, it
may cause a slight rise of A(Li) with metallicity. This expectation is
contradicted, however, by the apparent constancy of $^6$Li in halo stars.

Clearly, the 5\% $^6\mathrm{Li}$ observed in \object{HD~84937} and other halo
stars is not enough by itself to produce the slope with metallicity in the {\em
Spite plateau} found by \citet{ryan} and A06. To achieve this, the
$^6\mathrm{Li}$/$^7\mathrm{Li}$ ratio must be  higher at lower metallicities,
which is again contradicted by the observations. To reconcile the
absence of slope in the $^6$Li plateau with the slope in the $^7$Li plateau,
\citet{asplund} invoke a metallicity-dependent pre-main-sequence (PMS) depletion
of $^6$Li; after correction for this, their $^6$Li data show a slope similar to
that of $^7$Li. Although viable, this explanation appears somewhat contrived,
and the prediction of Li destruction is less certain on the PMS than on the main
sequence \citep{proffitt}.  For example, the models of \citet{piau05} are
marginally consistent with the $^6$Li measurements, but they do not lead to a slope
in $^6$Li.

\subsection{Pre-Galactic Li processing}

\citet{piau} suggested that the first massive supernovae have ejected large 
quantities of Li-poor (and D-poor) hydrogen. If these ejections are triggering
local star formation, the stars so formed will have inhomogeneous Li abundances.
The following yields of (more numerous) lower mass (less astration) supernovae
will provide slightly higher (and more homogeneous) Li abundances. The observed
Li abundance would be a reduced Li primordial abundance, especially for the most
metal-poor stars of the sample, only slightly reduced for the less metal-poor
ones. This scenario would not completely explain the gap
between the {\em Spite plateau} and the prediction derived from WMAP, but could
account for 0.2--0.3 dex.  \citet{piau} account for the remaining discrepancy by
appealing to depletion during the lifetimes of the low-mass stars.

A massive astration of Li in half of the Pop II material has been suggested by
\citet{piau}. This process destroys Li by a factor 2 in 2\% of the mass of the
Milky Way, i. e., the halo. Later the remaining 98\% of the mass of the Milky Way
joins the halo by infall. This infalling matter is made of primordial gas and
the mixture produces the Pop I. Is this large amount of primordial gas also
astrated depleting  Li? Not more than by a factor of 2, if the models of chemical
evolution of D in the Galaxy are to be trusted.
We note in passing that all the scenarios advocating the destruction of $^7$Li
imply an even larger destruction of $^6$Li.  On the other hand, production of
$^6$Li implies production of $^7$Li, thus some of the above scenarios may be
difficult to reconcile with the $^6$Li observations.

Finally, there is the possibility of a pre-Galactic origin for both isotopes as
suggested by \citet{asplund}. Possible $^6$Li production channels include
proto-galactic shocks \citep{suzuki,fields} and late-decaying or annihilating
supersymmetric particles during the era of big bang nucleosynthesis
\citep{Jedamzik3}, as mentioned below. The presence of $^6$Li limits the possible degree of stellar
$^7$Li depletion and thus sharpens the discrepancy with standard big bang
nucleosynthesis.

\subsection{New physics \label{NP}}

The dominant uncertainty in SBBN Li production is the cross section of the $\rm
^3He(\alpha,\gamma)^7Be$ reaction. This has been examined in detail by
\citet{cyburt}, who concluded that large errors in the cross section of this
reaction are unlikely. The possibility that $\rm ^7Be$ is destroyed by the
reactions $\rm ^7Be$ $(d,p)$ $\rm 2^4He$ and $^7$Be$(d,\alpha)^5$Li has been
considered by \citet{coc}, who concluded that the issue remains open, pending
accurately measured cross sections for these often-neglected reactions. However
\citet{angulo} have recently measured this cross section at energies appropriate
to the big bang environment and find it to be a factor of 10 {\em smaller} than
what was previously assumed. Thus, this possibility for reconciling the {\em Spite
plateau} with the baryonic density derived from the WMAP measurement no
longer appears viable. On the contrary, \citep{leonard} have recently provided a high
precision measurement of the $\rm ^2H({\mathit d, \mathit p})^3H$ and $\rm
^2H({\mathit d, \mathit n})^3He$ total cross sections, which imply an {\em
increase} of 0.02 dex of the predicted Li abundance, thus making the discrepancy
even larger.

Other possibilities exist, e.g., using non-standard BBN  to predict the
abundances of the light elements. The most appealing such model is the case of a
late-decaying massive particle \citep{Jedamzik1, Jedamzik2}. However, in the
case of an electromagnetic decay, this hypothesis implies either a D abundance
that is too low with respect to current observations, or a $\rm ^3He/D$ ratio
that is too high \citep{ellis}. Recently \citet{Jedamzik3} have shown that if
the late decaying particle is a gravitino, then the {\em Spite plateau} and the
baryonic density determined by WMAP may be reconciled. Even more interestingly,
a primordial production of $^6$Li at the level of what was observed in metal-poor
stars (\citealt{smith93,HT1994,cayrel99}; A06) can also be attained for
appropriate choices of the gravitino properties.

\subsection{What else?}

We find that none of the above scenarios is very satisfactory to describe
existing data as depicted by Fig.\ref{asplund}. However, the number of stars on
which these conclusions rest is still very small, and observations of more
extremely metal-poor stars are still needed to clarify the situation.

\section{Conclusions}

Our investigation has considerably increased the number of TO stars below
[Fe/H] = --3.0 with available $^7$Li abundance estimates. The interpretation of
the results is by no means straightforward. Our measurements alone do not
support the existence of any slope with metallicity or \teffns; however, an
increase in the dispersion at the lowest metallicities or an abrupt downturn of
Li abundances below [Fe/H] = --3.0 could be consistent with the data. When
considering the A06 data, which after suitable rescaling should be on
the same metallicity and temperature scale as our own data, sizeable slopes with
[Fe/H] and \teff are found. Our data are not in contradiction with the A06 
data, being somewhat complementary, in the sense that our data populate
the lowest metallicity regions. When considering the complete data set (A06
plus our own) the situation described by Fig. \ref{asplund} is that of a
slope with [Fe/H] and an increased scatter at the lowest metallicities. We stress
again that the slope is implied only by the data of A06,
relative to a sample showing a strong correlation
between [Fe/H] and \teff.

The existence of a {\em gap} between A(Li) in metal--poor stars and the
primordial Li predicted by the SBBN and the baryonic density determined by WMAP
is confirmed. The gap could be filled if metal--poor TO stars had effective
temperatures $\sim 7300$~K, which appears inconsistent with the colours of the
stars, the profiles of their H$\alpha$ lines, and their iron-excitation
equilibria. 
The temperature scale of metal-poor stars still has a zero point
uncertainty  at the level
of $\sim 200$ K. The iron-excitation equilibria in our stars do
not support extremely high temperature scales, such as that of
\citet{RM05}; however, our analysis does assume LTE and makes
use of 1D model atmospheres.  Departures from LTE and
granulation effects should be investigated to assess
the significance of our result.
In any case, the present observations constitute a strong constraint
for any theory seeking to explain the Li-WMAP discrepancy in terms of Li
depletion in metal-poor stars. Parallaxes for these stars would be extremely
valuable, since their TO status rests entirely on spectroscopic surface
gravities, which could be affected by systematic errors in the analysis (NLTE
effects on \ion{Fe}{i}, 3D effects, etc.); these should become available around 2020
from the GAIA mission and/or SIM. The issue of the temperature scale of
metal-poor stars is still not settled; a direct measurement of the angular
diameter of even one metal-poor star would be extremely valuable.

\begin{acknowledgements}

We are grateful to H.G. Ludwig for interesting discussions on the topic of 3D
radiative transfer, and for useful comments on an early version of this paper,
and also to S. Andrievsky for interesting discussions on the topic of NLTE. We
wish to thank M. Asplund for his help in our comparison with his results. PB
acknowledges support from the MIUR/PRIN 2004025729\_002 and from EU contract
MEXT-CT-2004-014265 (CIFIST). TCB acknowledges partial support from a series of
grants awarded by the US National Science Foundation, most recently, AST
04-06784, as well as from grant PHY 02-16783: Physics Frontier Center/Joint
Institute for Nuclear Astrophysics (JINA). BN and JA thank the Carlsberg
Foundation and the Danish Natural Science Research Council for financial
support. This research has made use of NASA's Astrophysics Data System. This
research has made use of the SIMBAD database, operated at CDS, Strasbourg,
France. This publication makes use of data products from the Two Micron All Sky
Survey, which is a joint project of the University of Massachusetts and the
Infrared Processing and Analysis Center/California Institute of Technology,
funded by the National Aeronautics and Space Administration and the National
Science Foundation.

\end{acknowledgements}

\bibliographystyle{aa}

\Online

{\scriptsize
\begin{longtable}{llllllllllllll}
\caption{\label{fe_1} Equivalent widths and abundances of iron lines}\\
\hline\hline\\
          &        &        &      & 
\multispan{2}{BS~16076-006\hfill}&\multispan{2}{CS~29527-015\hfill}&\multispan{2}{CS~22888-031\hfill}
&\multispan{2}{CS~22948-093\hfill}&\multispan{2}{CS~29518-043\hfill}\\
$\lambda$& $\chi$ & log gf & Ref. & EW  & A(Fe) & EW  & A(Fe) & EW  & A(Fe) & EW  & A(Fe) & EW  & A(Fe) \\
 nm      & eV     &        &      & pm  & dex   & pm  & dex   & pm  & dex   & pm  & dex   & pm  & dex   \\
\hline\hline\\
\endfirsthead \caption{continued.}\\ \hline\hline\\
          &        &        &      & 
\multispan{2}{BS~16076-006\hfill}&\multispan{2}{CS~29527-015\hfill}&\multispan{2}{CS~22888-031\hfill}
&\multispan{2}{CS~22948-093\hfill}&\multispan{2}{CS~29518-043\hfill}\\
$\lambda$& $\chi$ & log gf & Ref. & EW  & A(Fe) & EW  & A(Fe) & EW  & A(Fe) & EW  & A(Fe) & EW  & A(Fe) \\
 nm      & eV     &        &      & pm  & dex   & pm  & dex   & pm  & dex   & pm  & dex   & pm  & dex   \\
\hline\hline\\ \endhead \hline\hline \endfoot 
%
%
%
\ion{Fe}{i}\\
 337.0783  & 2.69 &$-0.266$  & BWL & \nod   & \nod   & \nod & \nod  & \nod & \nod & \nod & \nod & \nod &  \nod \\                   
 339.9333  & 2.20 &$ -0.622 $& BWL &   1.54 &  3.57  & \nod & \nod  & 1.38 & 4.38 & \nod & \nod & \nod & \nod  \\                   
 340.1519  & 0.92 &$ -2.059 $& BWL &   1.37 &  3.54  & \nod & \nod  & \nod & \nod & \nod & \nod & \nod & \nod  \\                   
 340.7460  & 2.18 &$ -0.020 $& BWL &   3.68 &  3.52  & 1.76 &  3.96 & 2.38 & 4.08 & \nod & \nod & 1.54 & 4.06  \\                   
 341.3132  & 2.20 &$ -0.404 $& BWL &   2.03 &  3.51  & \nod & \nod  & \nod &\nod  & \nod & \nod & \nod & \nod  \\                   
 341.7841  & 2.22 &$ -0.676 $& BWL &   \nod & \nod   &\nod  &\nod   &\nod  &\nod  & \nod & \nod & \nod & \nod  \\                   
 341.8507  & 2.22 &$ -0.761 $& BWL &   1.12 &  3.57  &\nod  &\nod   &\nod  &\nod  & \nod & \nod & \nod & \nod  \\
\hline\hline
\end{longtable}

\begin{longtable}{llllllllllllll}
\caption{\label{fe_2}Lines}\\
\hline\hline\\
          &        &        &      & 
\multispan{2}{CS~22177-009\hfill}&\multispan{2}{CS~30339-069\hfill}&\multispan{2}{CS~22966-011\hfill}
&\multispan{2}{BS~16968-061\hfill}&\multispan{2}{CS~22965-054\hfill}\\
$\lambda$& $\chi$ & log gf & Ref. & EW  & A(Fe) & EW  & A(Fe) & EW  & A(Fe) & EW  & A(Fe) & EW  & A(Fe) \\
 nm      & eV     &        &      & pm  & dex   & pm  & dex   & pm  & dex   & pm  & dex   & pm  & dex   \\
\hline\hline\\
\endfirsthead \caption{continued.}\\ \hline\hline\\
          &        &        &      & 
\multispan{2}{CS~22177-009\hfill}&\multispan{2}{CS~30339-069\hfill}&\multispan{2}{CS~22966-011\hfill}
&\multispan{2}{BS~16968-061\hfill}&\multispan{2}{CS~22965-054\hfill}\\
$\lambda$& $\chi$ & log gf & Ref. & EW  & A(Fe) & EW  & A(Fe) & EW  & A(Fe) & EW  & A(Fe) & EW  & A(Fe) \\
 nm      & eV     &        &      & pm  & dex   & pm  & dex   & pm  & dex   & pm  & dex   & pm  & dex   \\
\hline\hline\\ \endhead \hline\hline \endfoot 
\ion{Fe}{i}\\

337.0783 & 2.69 &$ -0.266$ & BWL   &   0.89 &  4.34 & \nod   & \nod  & \nod   & \nod   & \nod   & \nod   & \nod   & \nod   \\         
339.9333 & 2.20 &$ -0.622$ & BWL   &   0.77 &  4.17 & \nod   & \nod  &  1.12  &  4.31  & \nod   & \nod   & \nod   & \nod   \\         
340.1519 & 0.92 &$ -2.059$ & BWL   &  \nod  & \nod  & \nod   & \nod  &  0.60  &  4.25  & \nod   & \nod   & \nod   & \nod   \\         
340.7460 & 2.18 &$ -0.020$ & BWL   &   2.80 &  4.28 & \nod   & \nod  & \nod   & \nod   &  3.75  &  4.32  &  3.50  &  4.35  \\         
341.3132 & 2.20 &$ -0.404$ & BWL   &   1.34 &  4.23 & \nod   & \nod  &  1.73  &  4.32  &  1.89  &  4.25  &  1.64  &  4.23  \\         
341.7841 & 2.22 &$ -0.676$ & BWL   &   0.90 &  4.32 & \nod   & \nod  & \nod   & \nod   &  1.37  &  4.37  &  1.75  &  4.56  \\         
341.8507 & 2.22 &$ -0.761$ & BWL   &   0.69 &  4.28 & \nod   & \nod  & \nod   & \nod   &  1.48  &  4.49  & \nod   & \nod   \\
\end{longtable}

\begin{longtable}{llllllllllllll}
\caption{\label{fe_3}Lines}\\
\hline\hline\\
          &        &        &      & 
\multispan{2}{BS~16023-046\hfill}&\multispan{2}{BS~17570-063\hfill}&\multispan{2}{CS~29506-007\hfill}
&\multispan{2}{CS~22953-037\hfill}&\multispan{2}{CS~29506-090\hfill}\\
$\lambda$& $\chi$ & log gf & Ref. & EW  & A(Fe) & EW  & A(Fe) & EW  & A(Fe) & EW  & A(Fe) & EW  & A(Fe) \\
 nm      & eV     &        &      & pm  & dex   & pm  & dex   & pm  & dex   & pm  & dex   & pm  & dex   \\
\hline\hline\\
\endfirsthead \caption{continued.}\\ \hline\hline\\
          &        &        &      & 
\multispan{2}{BS~16023-046\hfill}&\multispan{2}{BS~17570-063\hfill}&\multispan{2}{CS~29506-007\hfill}
&\multispan{2}{CS~22953-037\hfill}&\multispan{2}{CS~29506-090\hfill}\\
$\lambda$& $\chi$ & log gf & Ref. & EW  & A(Fe) & EW  & A(Fe) & EW  & A(Fe) & EW  & A(Fe) & EW  & A(Fe) \\
 nm      & eV     &        &      & pm  & dex   & pm  & dex   & pm  & dex   & pm  & dex   & pm  & dex   \\
\hline\hline\\ \endhead \hline\hline \endfoot 
\ion{Fe}{i}\\                                                                                                                        
337.0783 & 2.69 &$-0.266$& BWL   &   0.84 &  4.39 & \nod   & \nod  & \nod   & \nod   & \nod   & \nod   &  1.48  &  4.64  \\         
339.9333 & 2.20 &$-0.622$& BWL   &  \nod  & \nod  &  2.18  &  4.75 & \nod   & \nod   &  1.68  &  4.67  &  1.99  &  4.72  \\         
340.1519 & 0.92 &$-2.059$& BWL   &   1.46 &  4.86 & \nod   & \nod  & \nod   & \nod   &  1.28  &  4.80  &  2.09  &  5.01  \\         
340.7460 & 2.18 &$-0.020$& BWL   &   2.86 &  4.38 &  2.86  &  4.33 &  3.31  &  4.40  &  2.91  &  4.40  &  3.65  &  4.53  \\         
341.3132 & 2.20 &$-0.404$& BWL   &  \nod  & \nod  & \nod   & \nod  & \nod   & \nod   &  1.76  &  4.48  &  2.09  &  4.53  \\         
341.7841 & 2.22 &$-0.676$& BWL   &  \nod  & \nod  &  1.48  &  4.58 & \nod   & \nod   & \nod   & \nod   &  1.81  &  4.74  \\         
341.8507 & 2.22 &$-0.761$& BWL   &  \nod  & \nod  & \nod   & \nod  & \nod   & \nod   & \nod   & \nod   &  1.72  &  4.79  \\
\end{longtable}

\begin{longtable}{llllllllllll}
\caption{\label{fe_4}Lines}\\
\hline\hline\\
          &        &        &      & 
\multispan{2}{CS~29518-020\hfill}&\multispan{2}{CS~30301-024\hfill}&\multispan{2}{CS~29499-060\hfill}
&\multispan{2}{CS~31061-032\hfill}\\
$\lambda$& $\chi$ & log gf & Ref. & EW  & A(Fe) & EW  & A(Fe) & EW  & A(Fe) & EW  & A(Fe) \\
 nm      & eV     &        &      & pm  & dex   & pm  & dex   & pm  & dex   & pm  & dex    \\
\hline\hline\\
\endfirsthead \caption{continued.}\\ \hline\hline\\
          &        &        &      & 
\multispan{2}{CS~29518-020\hfill}&\multispan{2}{CS~30301-024\hfill}&\multispan{2}{CS~29499-060\hfill}
&\multispan{2}{CS~31061-032\hfill}\\
$\lambda$& $\chi$ & log gf & Ref. & EW  & A(Fe) & EW  & A(Fe) & EW  & A(Fe) & EW  & A(Fe)  \\
 nm      & eV     &        &      & pm  & dex   & pm  & dex   & pm  & dex   & pm  & dex    \\
\hline\hline\\ \endhead \hline\hline \endfoot 
\ion{Fe}{i}\\ 
337.0783 & 2.69 &$-0.266$& BWL   &  \nod  & \nod  & \nod   & \nod  & \nod   & \nod   & \nod   & \nod    \\         
339.9333 & 2.20 &$-0.622$& BWL   &  \nod  & \nod  &  2.45  &  4.88 & \nod   & \nod   & \nod   & \nod    \\         
340.1519 & 0.92 &$-2.059$& BWL   &  \nod  & \nod  & \nod   & \nod  & \nod   & \nod   & \nod   & \nod    \\         
340.7460 & 2.18 &$-0.020$& BWL   &  \nod  & \nod  &  3.77  &  4.58 & \nod   & \nod   &  3.70  &  4.65   \\         
341.3132 & 2.20 &$-0.404$& BWL   &  \nod  & \nod  &  2.56  &  4.69 & \nod   & \nod   & \nod   & \nod    \\         
341.7841 & 2.22 &$-0.676$& BWL   &  \nod  & \nod  &  1.55  &  4.68 & \nod   & \nod   & \nod   & \nod    \\         
341.8507 & 2.22 &$-0.761$& BWL   &  \nod  & \nod  & \nod   & \nod  & \nod   & \nod   & \nod   & \nod    \\
\end{longtable}
\normalsize
\bibliographystyle{aa}

\begin{thebibliography}{}

\bibitem[Akritas \& Bershady(1996)]{ab} Akritas, M.~G.~\& 
Bershady, M.~A.\ 1996, \apj, 470, 706 

\bibitem[Alonso et al.(1994)]{Alonso94} Alonso, A., Arribas, S., 
\& Martinez-Roger, C.\ 1994, \aaps, 107, 365 

\bibitem[Alonso, Arribas, \& Mart{\'{\i}}nez-Roger(1996)]{alonso}
Alonso, A., Arribas, S., \& Mart{\'{\i}}nez-Roger, C.\ 1996, \aap, 313, 873  
(A96)

\bibitem[Alvarez \& Plez (1998)]{alvarez_plez} 
Alvarez R., Plez B., 1998, A\&A 330, 1109

\bibitem[Angulo et al.(2005)]{angulo} Angulo, C., et al.\ 
2005, \apjl, 630, L105 
 
\bibitem[Anstee et al.(1997)]{anstee} Anstee, S.~D., O'Mara, 
B.~J., \& Ross, J.~E.\ 1997, \mnras, 284, 202 

\bibitem[Aoki et al.(2004)]{aoki} Aoki, W., Inoue, S., 
Kawanomoto, S., Ryan, S.~G., Smith, I.~M., Suzuki, T.~K., \& Takada-Hidai, 
M.\ 2004, \aap, 428, 579 
 
\bibitem[Asplund et al. (1997)]{Asp97}Asplund, M., Gustafsson, B., Kiselman, D., Eriksson, K. 1997 A\&A 318, 521

\bibitem[Asplund et al.(2003)]{asp03} Asplund, M., Carlsson, 
M., \& Botnen, A.~V.\ 2003, \aap, 399, L31 

\bibitem[Asplund et al.(2006)]{asplund} Asplund, M., Lambert, 
D.~L., Nissen, P.~E., Primas, F., \& Smith, V.~V.\ 2006, \apj, 644, 229 

\bibitem[Balachandran (1990)]{Balachandran} Balachandran, S.\ 1990, 
\apj, 354, 310 

\bibitem[Barklem et al.(2000)]{barklem} Barklem, P.~S., 
Piskunov, N., \& O'Mara, B.~J.\ 2000, \aap, 363, 1091 
 
\bibitem[Barklem et al.(2002)]{barklem02} Barklem, P.~S., 
Stempels, H.~C., Allende Prieto, C., Kochukhov, O.~P., Piskunov, N., \& 
O'Mara, B.~J.\ 2002, \aap, 385, 951 
 
\bibitem[Beers (1999)]{beers99} Beers, T.~C.\ 1999, \apss, 265, 
547 

\bibitem[Beers et al.(1985)]{beers85} Beers, T.~C., Preston, 
G.~W., \& Shectman, S.~A.\ 1985, \aj, 90, 2089 
 
\bibitem[Beers et al.(1992)]{beers92} Beers, T.~C., Preston, 
G.~W., \& Shectman, S.~A.\ 1992, \aj, 103, 1987 

\bibitem[Bessell \& Brett (1988)]{BesselBrett} Bessell, M.~S.~\& Brett, J.~M.\ 1988, \pasp, 100, 1134

\bibitem[Boesgaard \& Tripicco(1986)]{BT} Boesgaard, 
A.~M., \& Tripicco, M.~J.\ 1986, \apjl, 302, L49 

\bibitem[Boesgaard et al.(2005)]{boesgaard} Boesgaard, A.~M., 
Stephens, A., \& Deliyannis, C.~P.\ 2005, \apj, 633, 398 
 
\bibitem[B{\"o}hm-Vitense(2004)]{erika} B{\"o}hm-Vitense, E.\ 
2004, \aj, 128, 2435 

\bibitem[Bonifacio(2002)]{B2002} Bonifacio, P.\ 2002,
Astrophysics and Space Science Library  
Vol.~274, Dordrecht: Kluwer Academic Publishers:  
New Quests in Stellar Astrophysics: the Link Between Stars and 
Cosmology, M. Ch\`avez, A. Bressan, A. Buzzoni \& D. Mayya eds. p. 77
 
\bibitem[{Bonifacio \& Molaro (1997)}]{BM97}
Bonifacio, P. \& Molaro, P. 1997, MNRAS, 285, 847

\bibitem[Bonifacio et al.(2000a)]{BMB00} Bonifacio, P., Monai, 
S., \& Beers, T.~C.\ 2000, \aj, 120, 2065 
 
\bibitem[Bonifacio et al.(2000b)]{BCM} Bonifacio, P., 
Caffau, E., \& Molaro, P.\ 2000, \aaps, 145, 473 
 
\bibitem[Bonifacio et al.(2002)]{B02} Bonifacio, P., et 
al.\ 2002, \aap, 390, 91 
 
\bibitem[Bonifacio et al.(2003)]{BIAU} Bonifacio, P., et 
al.\ 2003, Elemental Abundances in Old Stars and Damped Lyman-{$\alpha$} 
Systems, 25th meeting of the IAU, Joint Discussion 15, 22 July 2003, 
Sydney, Australia, 15,  

\bibitem[Castilho et al.(2000)]{castilho} Castilho, B.~V., 
Pasquini, L., Allen, D.~M., Barbuy, B., \& Molaro, P.\ 2000, \aap, 361, 92 
 
\bibitem[Cayrel(1988)]{cayrel88} Cayrel, R.\ 1988, IAU 
Symp.~132: The Impact of Very High S/N Spectroscopy on Stellar Physics, 
132, 345 
 
\bibitem[Cayrel et al.(1999)]{cayrel99} Cayrel, R., Spite, M., 
Spite, F., Vangioni-Flam, E., Cass{\' e}, M., \& Audouze, J.\ 1999, \aap, 
343, 923 

\bibitem[Cayrel \& Steffen(2000)]{CS00} Cayrel, R., \& 
Steffen, M.\ 2000, IAU Symposium, 198, 437 

\bibitem[Cayrel et al.(2004)]{cayrel04} Cayrel, R., et al.\ 
2004, \aap, 416, 1117 

\bibitem[Charbonnel \& Primas(2005)]{CP05} Charbonnel, C., 
\& Primas, F.\ 2005, \aap, 442, 961 

\bibitem[Coc et al.(2004)]{coc} Coc, A., Vangioni-Flam, E., 
Descouvemont, P., Adahchour, A., \& Angulo, C.\ 2004, \apj, 600, 544 

\bibitem[Collet et al.(2006)]{Collet} Collet, R., Asplund, M., 
\& Trampedach, R.\ 2006, ArXiv Astrophysics e-prints, 
arXiv:astro-ph/0605219 

\bibitem[Cutri et al. (2003)]{cutri} Cutri, R.M. et al.\ 2003
Explanatory Supplement to the 2MASS All Sky Data Release,
http://www.ipac.caltech.edu/2mass/releases/allsky
/doc/explsup.html

\bibitem[Cyburt et al.(2004)]{cyburt} Cyburt, R.~H., Fields, 
B.~D., \& Olive, K.~A.\ 2004, \prd, 69, 123519 
 
\bibitem[Dearborn et al.(1992)]{dearborn} Dearborn, D.~S.~P., 
Schramm, D.~N., \& Hobbs, L.~M.\ 1992, \apjl, 394, L61 

\bibitem[Dekker et al.(2000)]{dekker} Dekker, H., D'Odorico, 
S., Kaufer, A., Delabre, B., \& Kotzlowski, H.\ 2000, \procspie, 4008, 534 

\bibitem[Deliyannis, Demarque, \& Kawaler(1990)]{deli} 
Deliyannis, C.~P., Demarque, P., \& Kawaler, S.~D.\ 1990, \apjs, 73, 21 

\bibitem[Deliyannis \& Ryan(2000)]{DR2000} Deliyannis, C.~P., 
\& Ryan, S.~G.\ 2000, Bulletin of the American Astronomical Society, 32, 
684 
 
\bibitem[Edvardsson et al. (1993)]{Edv93}Edvardsson, B.,
Andersen, J., Gustafsson, B., Lambert, D.L., Nissen, P.E., Tomkin, J. 1993 A\& A 275, 101

\bibitem[Ellis et al.(2005)]{ellis} Ellis, J., Olive, K.~A., 
\& Vangioni, E.\ 2005, Physics Letters B, 619, 30 
 
\bibitem[Fields \& Prodanovi{\'c}(2005)]{fields} Fields, 
B.~D., \& Prodanovi{\'c}, T.\ 2005, \apj, 623, 877 

\bibitem[Fran\c cois et al. (2003)]{francois}  Fran\c cois P.,
Depagne E.,  Hill V., Spite M., Spite F., et al. \ 2003, \aap, 403, 1105 

\bibitem[Frebel et al. (2005)]{frebel} Frebel, A., et al. 2005, Nature, 434, 871

\bibitem[Fuhrmann et al.(1993)]{fuhrmann}Fuhrmann, K., Axer, 
M., \& Gehren, T.\ 1993, \aap, 271, 451 


\bibitem[Garc{\'{\i}}a P{\'e}rez et al.(2006)]{ana} 
Garc{\'{\i}}a P{\'e}rez, A.~E., Asplund, M., Primas, F., Nissen, P.~E., \& 
Gustafsson, B.\ 2006, \aap, 451, 621 

\bibitem[Glaspey, Pritchet, \& Stetson(1994)]{glaspey} Glaspey, 
J.~W., Pritchet, C.~J., \& Stetson, P.~B.\ 1994, \aj, 108, 271 

\bibitem[Gratton et al.(2001)]{gratton} Gratton, R.~G., et al.\ 
2001, \aap, 369, 87 

\bibitem[Gustafsson et al. (1975)]{Gus75} Gustafsson, B., Bell, R.A., Eriksson, K., Nordlund \AA ., 1975, A\&A 42, 407

\bibitem[Gustafsson et al. (2003)]{GEE03}
Gustafsson B.,  Edvardsson B., Eriksson K., Graae-J{\o}rgensen U.,
Mizuno-Wiedner, M., Plez, B., 2003, in Stellar Atmosphere Modeling, 
eds. I. Hubeny, D. Mihalas, K. Werner, ASP Conf. Series 288, 331. 

\bibitem[Hobbs \& Thorburn(1994)]{HT1994} Hobbs, L.~M.~\& 
Thorburn, J.~A.\ 1994, \apjl, 428, L25 

\bibitem[Jedamzik(2004a)]{Jedamzik1} Jedamzik, K.\ 2004, \prd, 70, 
083510 
 
\bibitem[Jedamzik(2004b)]{Jedamzik2} Jedamzik, K.\ 2004, \prd, 70, 
063524 

\bibitem[Jedamzik et al.(2006)]{Jedamzik3} Jedamzik, K., Choi, 
K.-Y., Roszkowski, L., \& Ruiz de Austri, R.\ 2006, Journal of Cosmology 
and Astro-Particle Physics, 7, 7 
 
\bibitem[Kiselman(1997)]{Kisel97} Kiselman, D.\ 1997, \apjl, 
489, L107 

\bibitem[Kiselman(1998)]{Kisel98} Kiselman, D.\ 1998, \aap, 
333, 732 

\bibitem[Korn et al. (2006)]{Korn06} Korn, A.J., Grundahl, F., 
Richard, O. et al. \ 2006 \nat 442, 657

\bibitem[Lambert \& Reddy(2004)]{lambertreddy} Lambert, D.~L., \& 
Reddy, B.~E.\ 2004, \mnras, 349, 757 

\bibitem[Leonard et al.(2006)]{leonard} Leonard, D.~S., 
Karwowski, H.~J., Brune, C.~R., Fisher, B.~M., \& Ludwig, E.~J.\ 2006, 
\prc, 73, 045801 
 
\bibitem[Mel{\' e}ndez \& Ram{\'{\i}}rez(2004)]{melendez} 
Mel{\' e}ndez, J., \& Ram{\'{\i}}rez, I.\ 2004, \apjl, 615, L33 

\bibitem[Mel{\'e}ndez et al.(2006)]{RM06} Mel{\'e}ndez, J., 
Shchukina, N.~G., Vasiljeva, I.~E., \& Ram{\'{\i}}rez, I.\ 2006, \apj, 642, 
1082 

\bibitem[Michaud(1986)]{michaud86} Michaud, G.\ 1986, \apj, 302, 
650 

\bibitem[Michaud, Richard, \& Richer(2004)]{michaud} Michaud, 
G., Richard, O., \& Richer, J.\ 2004, Memorie della Societ\`a Astronomica 
Italiana, 75, 339 

\bibitem[Nissen et al.(2000)]{nissen} Nissen, P.~E., Asplund, 
M., Hill, V., \& D'Odorico, S.\ 2000, \aap, 357, L49 

\bibitem[Norris et al.(1997)]{norris97} Norris, J.~E., Ryan, 
S.~G., Beers, T.~C., \& Deliyannis, C.~P.\ 1997, \apj, 485, 370 
 
\bibitem[Norris et al.(2000)]{norris00} Norris, J.~E., Beers, 
T.~C., \& Ryan, S.~G.\ 2000, \apj, 540, 456 

\bibitem[Plez et al.(1992)]{Ple92} Plez, B., Brett, J.M., Nordlund, \AA.1992  A\&A 256,551

\bibitem[Piau(2005)]{piau05} Piau, L., ArXiv Astrophysics e-prints, arXiv:astro-ph/0511402 
 
\bibitem[Piau et al.(2006)]{piau} Piau, L.; Beers, T. C., Balsara, D. S.,
Sivarani, T., Truran, J. W., \& Ferguson, J. W. 2006, \apj, in press,
ArXiv Astrophysics e-prints, arXiv:astro-ph/0603553 
 
\bibitem[Pinsonneault et al.(1992)]{pin92} Pinsonneault, 
M.~H., Deliyannis, C.~P., \& Demarque, P.\ 1992, \apjs, 78, 79 

\bibitem[Press et al. (1992)]{nr} Press, W.~H., Teukolsky, S.~A., 
Vetterling, W.~T., \& Flannery, B.~P.\ 1992, Cambridge: University Press, 
1992, 2nd ed.,  

\bibitem[Proffitt \& Michaud(1989)]{proffitt} Proffitt, C.~R., 
\& Michaud, G.\ 1989, \apj, 346, 976 

\bibitem[Proffitt et al.(1990)]{proffitt1990} Proffitt, C.~R., 
Michaud, G., \& Richer, J.\ 1990, ASP Conf.~Ser.~  9: Cool Stars, Stellar 
Systems, and the Sun, 9, 351 

\bibitem[Ram{\'{\i}}rez \& Mel{\'e}ndez(2005)]{RM05} 
Ram{\'{\i}}rez, I., \& Mel{\'e}ndez, J.\ 2005, \apj, 626, 465 
(RM05)

\bibitem[Richard, Michaud, \& Richer(2002)]{ric3} Richard, 
O., Michaud, G., \& Richer, J.\ 2002, \apj, 580, 1100 

\bibitem[Richard et al.(2002)]{ric1} Richard, O., Michaud, 
G., Richer, J., Turcotte, S., Turck-Chi{\` e}ze, S., \& VandenBerg, D.~A.\ 
2002, \apj, 568, 979 

\bibitem[Richard et al.(2005)]{richard} Richard, O., Michaud, G., Richer, J.
2005, \apj, 619, 538
                                                                                
\bibitem[Ryan et al. (1996)]{ryan96} 
Ryan, S.~G., Beers, T.~C., Deliyannis, C.~P., \& Thorburn, J.~A.\ 1996, 
\apj, 458, 543 

\bibitem[Ryan \& Deliyannis(1998)]{ryan98} Ryan, S.~G., \& 
Deliyannis, C.~P.\ 1998, \apj, 500, 398 
 
\bibitem[Ryan, Norris, \& Beers(1999)]{ryan} Ryan, S.~G., 
Norris, J.~E., \& Beers, T.~C.\ 1999, \apj, 523, 654 

\bibitem[Ryan et al.(2000)]{ryan00} Ryan, S.~G., Beers, T.~C., 
Olive, K.~A., Fields, B.~D., \& Norris, J.~E.\ 2000, \apjl, 530, L57 

\bibitem[Ryan et al.(2001)]{ryan01} Ryan, S.~G., Beers, T.~C., 
Kajino, T., \& Rosolankova, K.\ 2001, \apj, 547, 231 

\bibitem[Ryan et al.(2002)]{ryan02} Ryan, S.~G., Gregory, 
S.~G., Kolb, U., Beers, T.~C., \& Kajino, T.\ 2002, \apj, 571, 501 

\bibitem[Sivarani et al. (2006)]{siva06} Sivarani, T.,
Beers, T.C., Bonifacio, P., Molaro, P., Cayrel, R., et al.
2006, \aap, in press, ArXiv Astrophysics e-prints, arXiv:astro-ph/0608112  

\bibitem[Schlegel et al.(1998)]{schlegel} Schlegel, D.~J., 
Finkbeiner, D.~P., \& Davis, M.\ 1998, \apj, 500, 525 
 
\bibitem[Schramm et al.(1990)]{SSD} Schramm, D.~N., 
Steigman, G., \& Dearborn, D.~S.~P.\ 1990, \apjl, 359, L55 

\bibitem[Smith, Lambert, \& Nissen(1993)]{smith93} Smith, 
V.~V., Lambert, D.~L., \& Nissen, P.~E.\ 1993, \apj, 408, 262 

\bibitem[Smith et al.(1998)]{smith98} Smith, V.~V., Lambert, 
D.~L., \& Nissen, P.~E.\ 1998, \apj, 506, 405 
 
\bibitem[Spergel et al.(2003)]{spergel} Spergel, D.~N., et al.\ 
2003, \apjs, 148, 175 

\bibitem[Spergel et al.(2006)]{spergel06} Spergel, D.~N., et al.\ 
2006, ArXiv Astrophysics e-prints, arXiv:astro-ph/0603449 

\bibitem[Spite \& Spite(1982a)]{spite82} Spite, M.~\& Spite, F.\ 
1982, \nat, 297, 483 

\bibitem[Spite \& Spite(1982b)]{ss82A} Spite, F., \& Spite, 
M.\ 1982, \aap, 115, 357 
 
\bibitem[Spite et al.(1996)]{spite96} Spite, M., Francois, P., 
Nissen, P.~E., \& Spite, F.\ 1996, \aap, 307, 172 

\bibitem[Spite et al.(2000)]{spite_natal} Spite, M., Spite, F., 
Cayrel, R., Hill, V., Depagne, E., Nordstr{\" o}m, B., \& Beers, T.~C.\ 
2000, IAU Symposium, 198, 356 

\bibitem[Spite et al.(2005)]{iaus05} Spite, M., et al.\ 2005, 
IAU Symposium, 228, 185 
 
\bibitem[Suzuki \& Inoue(2002)]{suzuki} Suzuki, T.~K., \& 
Inoue, S.\ 2002, \apj, 573, 168 

\bibitem[Talon \& Charbonnel(2003)]{talon} Talon, S., \& 
Charbonnel, C.\ 2003, \aap, 405, 1025 

\bibitem[Thorburn(1994)]{thorburn} Thorburn, J.~A.\ 1994, \apj, 
421, 318 

\bibitem[VandenBerg \& Clem(2003)]{clem} VandenBerg, D.~A., 
\& Clem, J.~L.\ 2003, \aj, 126, 778 
 
\bibitem[Vauclair \& Charbonnel(1995)]{VC95} Vauclair, S., 
\& Charbonnel, C.\ 1995, \aap, 295, 715 
 
\bibitem[van't Veer-Menneret \& M\'egessier(1996)]{vantveer} van't 
Veer-Menneret, C., \& M\'egessier, C.\ 1996, \aap, 309, 879 

\bibitem[Vidal et al.(1973)]{vcs} Vidal, C.~R., Cooper, J., 
\& Smith, E.~W.\ 1973, \apjs, 25, 37 
 
\bibitem[Wagoner et al.(1967)]{wagoner} Wagoner, R.~V., Fowler, 
W.~A., \& Hoyle, F.\ 1967, \apj, 148, 3 

\end{thebibliography}

\begin{thebibliography}{}
\bibitem[Bard et al.(1991)]{1991A&A...248..315B} Bard, A., Kock, A., \& 
Kock, M.\ 1991, \aap, 248, 315 (BKK)
\bibitem[Bridges(1973)]{1973pig..conf..418B} Bridges, J.~M.\ 1973, 
Phenomena in Ionized Gases, Eleventh International Conference, 418 (B)
\bibitem[Fuhr, Martin \& Wiese]{FMW}  Fuhr, J.R., Martin, G.A., and Wiese, W.L. \ 1988
Journal of Physical and Chemical Reference Data, 17, Suppl. 4 (FMW)
\bibitem[O'Brian et al.(1991)]{1991JOSAB...8.1185O} O'Brian, T.~R., 
Wickliffe, M.~E., Lawler, J.~E., Whaling, W., \& Brault, J.~W.\ 1991, 
Journal of the Optical Society of America B: Optical Physics, Volume 8, 
Issue 6, June 1991, pp.1185-1201, 8, 1185 (BWL)
\end{thebibliography}
{}

\begin{longtable}{llllllllr}
\caption{\label{logobs}Log of observations}\\
\hline\hline
Star & $\alpha(2000)$ & $\delta(2000)$ &Date & $t_{exp}$(s) & $\lambda_c$ (nm)  & MJD-24000.5& $v_R$ (\kms )  \\

\hline\hline
\endfirsthead \caption{continued.}\\ \hline\hline

Star & $\alpha(2000)$ & $\delta(2000)$ &Date & $t_{exp}$(s) & $\lambda_c$ (nm)  & MJD-24000.5& $v_R$ (\kms )  \\
\hline \endhead \hline \endfoot 

\object{ BS 16076-006} & 12:48:22.7& +20:56:41& 2001-06-02  &3600 &  573    &   52062.040054400 &  206.0 \\
                       &           &          & 2001-06-02  & 3600& 396     &   52062.039607600 &  206.0 \\
                       &           &          & 2001-06-02  &3600 &  850    &   52062.083546400 &  208.0 \\ 
                       &           &          & 2001-06-02  & 3600& 396     &   52062.083098100 &  205.8 \\
                       &           &          & 2001-06-03  &3600 & 573     &   52063.957226600 &  205.2 \\
                       &           &          & 2001-06-03  & 3600& 396     &   52063.956792500 &  205.2 \\
\object{ CS 29527-015} & 00:29:10.6&-19:10:06 & 2000-10-16  &3600 & 573     &   51833.126082100 &  51.4 \\
                       &           &          & 2001-09-06  &3600 & 396     &   52158.257443400 &  50.7 \\
\end{longtable}

\appendix

\section{Comparison of temperature and metallicity scales}

\normalsize

\begin{figure}
\resizebox{\hsize}{!}{\includegraphics[clip=true]{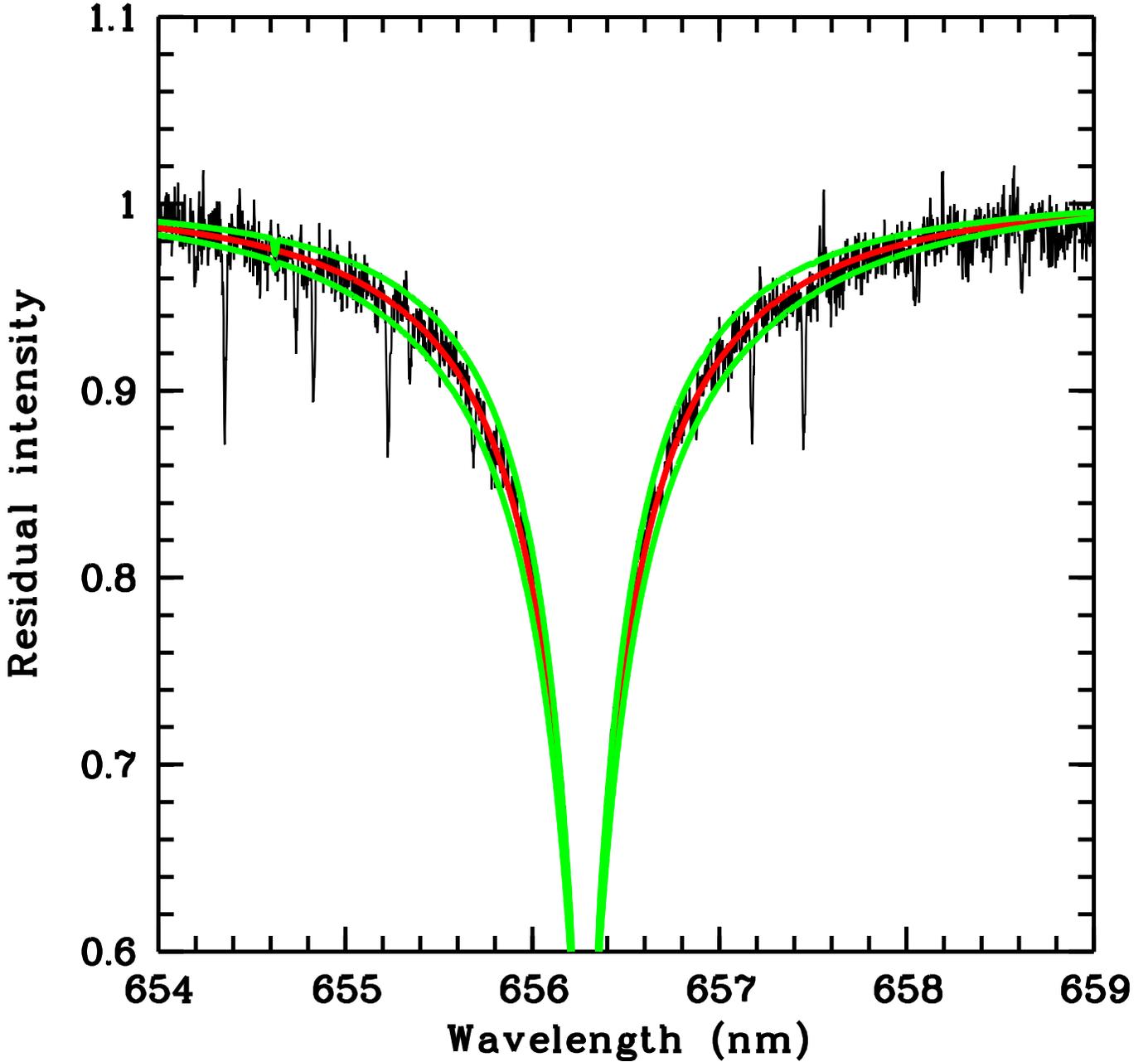}}
\caption{
Example of fits to the H$\alpha$ line of the star BS~16023-043. The best-fit profile
corresponds to \teff = 6364~K.  The other two profiles shown correspond
to \teff $\pm 200$~K of this value. The narrow absorption features
are $\mathrm{H_2O}$ telluric lines.
}
\label{HA_bs16023_043}
\end{figure}

This appendix is devoted to a detailed comparison
of the temperature and metallicity scales of
\citet[][, hereafter A06]{asplund} and the ones derived in 
the present paper.  Since there are no stars in common
among the two studies we downloaded UVES spectra for two of the
most metal-poor stars of A06 from the
ESO archive: \object{LP 815-43}
and \object{CD --33 1173}. The list of
spectra is given in Table \ref{archivedata} and includes
all of the data used by A06, taken
with the image slicer,  as well as other data 
taken with the slit and more similar to our LP data.
The purpose of this exercise is to compare
data reduction and analysis procedures.
An example of an H$\alpha$ fit for
one of our program stars is given in Fig. \ref{HA_bs16023_043}.

\begin{figure}
\resizebox{\hsize}{!}{\includegraphics[clip=true]{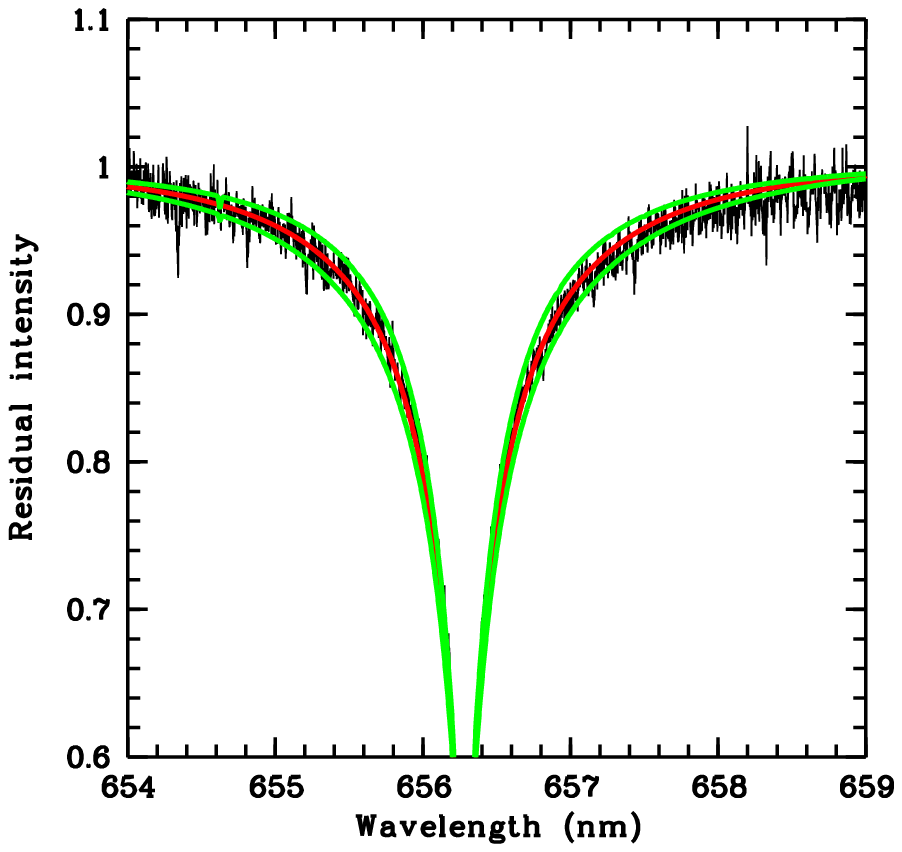}}
\caption{
The H$\alpha$ profile of  LP~815-043.  The upper profile is
from the 2004 slicer data, while the lower profile is from the
2001 slit data.  It is clear that the slicer profile is distorted.
}
\label{HA_LP815_43_slic_slit}
\end{figure}

Our data reduction procedure includes division by a normalized
flat-field, which achieves a correction of both the
detector pixel-to-pixel variations and the echelle blaze function.
The blaze function pattern is removed at the level of a few
percent in our slit spectra, which is sufficient to
obtain reliable profile of H$\alpha$, as shown in Fig. \ref{HA_bs16023_043}.
The procedure fails on the data taken with image slicer \# 3,
as shown in Fig. \ref{HA_LP815_43_slic_slit}, which compares
a spectrum of star LP 815-43 taken with the image slicer with
another spectrum taken without the image slicer.
The blaze profile of the stellar spectrum (taken through
the image slicer) appears significantly different from
that of the flat-field spectrum (which is taken without the slicer,
through a slit).
The reduction procedure adopted by A06 differs from ours 
in that the blaze function is approximated by a polynomial fit
directly on the stellar spectrum (the spectra of metal-poor stars
are line-free enough to make this procedure unambiguous), and
for the order containing H$\alpha$ for which this is impossible,
the blaze profile is obtained by interpolating between the
profiles of the two adjacent orders.
Clearly, if we adopted the same procedure we would obtain
results identical, or at least very similar ones to those
of A06; however, this would not tell us
much about the similarity of the two temperature scales.

\begin{figure}
\resizebox{\hsize}{!}{\includegraphics[clip=true]{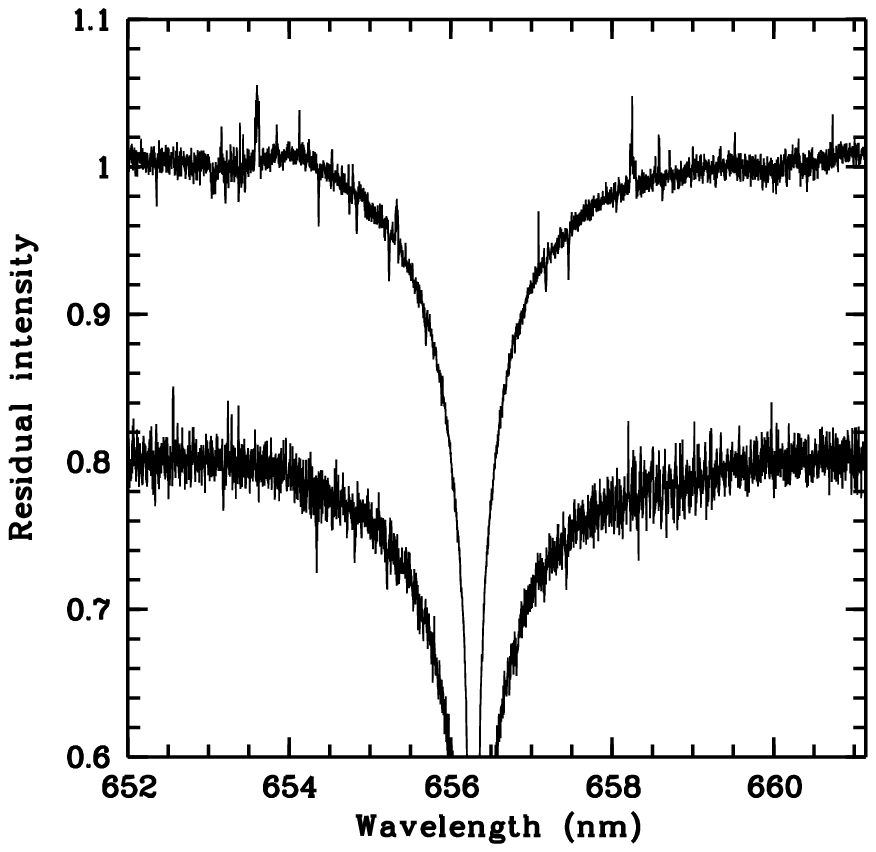}}
\caption{
Example of fits to the H$\alpha$ line of the star LP 815-043, based on the
2001 slit data. The best-fit profile
corresponds to \teff = 6409~K, while the other two profiles shown correspond
to \teff $\pm 200$~K of this value.
}
\label{HA_LP815_43_slit2}
\end{figure}

Instead we preferred to use the slit data that  was available for the
two stars and that is very similar to our LP data.
For star LP 815-43, from one spectrum we obtain \teff = 6364 K
and for the other \teff = 6409 K, the mean is 6387 K,
to be compared with 6400 K of A06 i.e. 13 K cooler.
For CD --33 1173 we have 6 spectra, the mean \teff is 6326 K with 
a standard deviation of 34 K, to be compared with \teff = 6390 K 
of A06, i.e., 64 K cooler.
The average of these two differences is $\sim 38$K.
This is considerably smaller than our estimate of
the error on \teff. 
Therefore our interpretation is that our temperatures and
those of A06 are on the same scale,
although a zero point shift of up to $\sim 40$K is possible.

The comparison of the metallicities scale requires more effort
and was therefore done only for LP 815-43.
We used the spectra in Table \ref{archivedata}
to measure the equivalent widths of the \ion{Fe}{i}
and  \ion{Fe}{ii} lines that have been used for all
of our programme stars. 
These have been used, together with our model atmospheres with
the \teff = 6400, as input to the  {\tt turbospectrum} code
to determine the abundances.
The surface gravity and microturbulent velocity were
iteratively adjusted to attain iron ionization equilibrium
and independence of abundance from equivalent width.
Our final parameters for this star are
[Fe/H]=-2.94 log g = 3.90 and $\xi = 1.61$\kms.
The line by line 
results are given  in Table \ref{fe_LP815_43}.
We note that our gravity is consistent, within errors
with the gravity of A06 (log g = 4.17),
which is derived from Str\"omgren photometry for this
star and therefore has an error of $\sim 0.25$dex,
like our own.
We therefore conclude that the A06 metallicity 
scale over-estimates [Fe/H] by 0.2 dex and apply this rigid shift to
all of his data.

{\scriptsize
\begin{longtable}{llllll}
\caption{\label{archivedata} Archive data for LP 815-43 and CD --33 1173}\\
\hline\hline\\
Star & Date & $t_{exp}$(s) & $\lambda_c$ (nm) & Slit  & MJD-24000.5   \\
\hline\hline\\
\endfirsthead\hline\hline \caption{continued.}\\ \hline\hline\\
Star & Date & $t_{exp}$(s) & $\lambda_c$ (nm) & Slit  & MJD-24000.5   \\
\hline\hline\\ \endhead \hline\hline \endfoot 
LP 815-43 & 1999-10-16 & 5400 & 346 & 0\farcs{8} & 51467.018148000 \\
          & 1999-10-16 & 2700 & 346 & 1\farcs{0} & 51459.008923800 \\
          & 2000-07-24 & 1800 & 705 & SLIC\# 3 & 51749.164882190 \\
          & 2000-07-24 & 1800 & 705 & SLIC\# 3 & 51749.186416570 \\
          & 2000-07-25 & 1200 & 705 & SLIC\# 3 & 51750.211278810 \\
          & 2000-07-25 & 1200 & 705 & SLIC\# 3 & 51750.225785770 \\
          & 2001-06-06 & 1200 & 437 & 0\farcs{7} & 52066.366975750 \\
          & 2001-06-06 & 1200 & 437 & 0\farcs{7} & 52066.381358710 \\
          & 2001-06-06 & 1200 & 437 & 0\farcs{7} & 52066.395742820 \\
          & 2001-11-26 &  300 & 580 & 0\farcs{8} & 52239.029463940 \\
          & 2001-11-26 &  600 & 580 & 0\farcs{8} & 52240.015000550 \\
          & 2004-08-30 & 3000 & 600 &  SLIC\# 3  & 53247.147858160 \\
          & 2004-08-30 & 3000 & 600 &  SLIC\# 3  & 53247.185733820 \\
          & 2004-08-30 & 3000 & 600 &  SLIC\# 3  & 53247.222292000 \\
          & 2004-08-31 & 3000 & 600 &  SLIC\# 3  & 53248.011494530 \\
          & 2004-08-31 & 3000 & 600 &  SLIC\# 3  & 53248.048641650 \\
          & 2004-08-31 & 3000 & 600 &  SLIC\# 3  & 53248.184532180 \\
CD --33 1173 & 2002-01-12 & 1500 & 580 & 0\farcs{9} & 52286.045541490 \\         
             & 2002-01-12 & 1500 & 580 & 0\farcs{9} & 52286.063493190 \\         
             & 2002-01-12 & 1500 & 580 & 0\farcs{9} & 52286.081448180 \\         
             & 2002-01-12 & 1500 & 580 & 0\farcs{9} & 52286.102812080 \\         
             & 2002-01-12 & 1500 & 580 & 0\farcs{9} & 52286.120760860 \\         
             & 2002-01-12 & 1500 & 580 & 0\farcs{9} & 52286.138718950 \\         
             & 2002-02-06 & 1800 & 705 & SLIC\# 3   & 52311.017325910 \\         
             & 2002-02-06 & 1800 & 705 & SLIC\# 3   & 52311.039937800 \\         
             & 2002-02-06 & 1800 & 705 & SLIC\# 3   & 52311.062072870 \\         
\hline\hline\\
\end{longtable}
}

{\scriptsize
\begin{longtable}{llllll}
\caption{\label{fe_LP815_43} Equivalent widths  of iron lines and abundances}\\
\hline\hline\\
         &        &        &      &\multispan{2}{LP 815-43\hfill}\\
$\lambda$& $\chi$ & log gf & Ref. & EW  & A(Fe) \\
 nm      & eV     &        &      & pm  & dex   \\
\hline\hline\\
\endfirsthead\caption{continued.}\\ \hline\hline\\
         &        &        &      &\multispan{2}{LP 815-43\hfill}\\
$\lambda$& $\chi$ & log gf & Ref. & EW  & A(Fe) \\
 nm      & eV     &        &      & pm  & dex   \\
\hline\hline\\ \endhead \hline\hline \endfoot 
\ion{Fe}{i}                  \\                                                                                                      
335.5228 & 3.30 &$-0.405$& BWL   &   0.28 &  4.62         \\         
337.0783 & 2.69 &$-0.266$& BWL   &   0.86 &  4.46         \\         
339.9333 & 2.20 &$-0.622$& BWL   &   1.18 &  4.53         \\         
340.1519 & 0.92 &$-2.059$& BWL   &   0.48 &  4.37         \\         
340.7460 & 2.18 &$-0.020$& BWL   &   2.95 &  4.46         \\         
341.3132 & 2.20 &$-0.404$& BWL   &   1.66 &  4.49         \\         
341.7841 & 2.22 &$-0.676$& BWL   &   1.11 &  4.57         \\         
341.8507 & 2.22 &$-0.761$& BWL   &   0.81 &  4.50         \\         
342.4284 & 2.18 &$-0.703$& BWL   &   0.98 &  4.49         \\         
342.5010 & 3.05 &$-0.500$& BWL   &   0.38 &  4.62         \\         
342.6383 & 0.99 &$-1.909$& BWL   &   0.68 &  4.45         \\         
342.7119 & 2.18 &$-0.098$& BWL   &   2.90 &  4.52         \\         
342.8193 & 2.20 &$-0.822$& BWL   &   0.77 &  4.51         \\         
344.0606 & 0.00 &$-0.673$& BWL   &   8.13 &  4.67         \\         
344.0989 & 0.05 &$-0.958$& BWL   &   7.13 &  4.66         \\         
344.3876 & 0.09 &$-1.374$& BWL   &   5.45 &  4.55         \\         
344.5149 & 2.20 &$-0.535$& BWL   &   1.30 &  4.49         \\         
344.7278 & 2.20 &$-1.021$& BWL   &   0.40 &  4.40         \\         
345.0328 & 2.22 &$+0.902$& BWL   &   0.52 &  4.43         \\         
345.2275 & 0.96 &$-1.919$& BWL   &   0.89 &  4.56         \\         
347.5450 & 0.09 &$-1.054$& BWL   &   6.73 &  4.64         \\         
347.6702 & 0.12 &$-1.507$& BWL   &   5.00 &  4.59         \\         
348.5340 & 2.20 &$-1.149$& BWL   &   0.40 &  4.53         \\         
349.0574 & 0.05 &$-1.105$& BWL   &   6.67 &  4.64         \\         
349.7841 & 0.11 &$-1.549$& BWL   &   4.93 &  4.60         \\         
352.1261 & 0.92 &$-0.988$& BWL   &   3.96 &  4.52         \\         
353.3198 & 2.88 &$-0.112$& BWL   &   1.65 &  4.80         \\         
353.6556 & 2.88 &$+0.115$& BWL   &   1.42 &  4.49         \\         
354.1083 & 2.85 &$+0.252$& BWL   &   1.75 &  4.45         \\         
354.2076 & 2.87 &$+0.207$& BWL   &   1.72 &  4.49         \\         
355.3739 & 3.57 &$+0.269$& BWL   &   0.68 &  4.59         \\         
355.4118 & 0.96 &$-2.206$& BWL   &   0.42 &  4.48         \\         
355.4925 & 2.83 &$+0.538$& BWL   &   2.75 &  4.43         \\         
355.6878 & 2.85 &$-0.040$& FMW   &   2.04 &  4.83         \\         
356.5379 & 0.96 &$-0.133$& BWL   &   6.80 &  4.51         \\         
358.1193 & 0.86 &$+0.406$& FMW   &   9.22 &  4.62         \\         
358.4659 & 2.69 &$-0.157$& BWL   &   1.82 &  4.73         \\         
358.5319 & 0.96 &$-0.802$& BWL   &   5.20 &  4.69         \\         
358.5705 & 0.92 &$-1.187$& FMW   &   3.49 &  4.60         \\         
358.6113 & 3.24 &$+0.173$& BWL   &   1.08 &  4.61         \\         
358.9105 & 0.86 &$-2.115$& FMW   &   1.04 &  4.73         \\         
360.3204 & 2.69 &$-0.256$& BWL   &   0.99 &  4.51         \\         
360.6679 & 2.69 &$+0.323$& BWL   &   2.73 &  4.51         \\         
360.8859 & 1.01 &$-0.100$& FMW   &   6.80 &  4.52         \\         
361.0159 & 2.81 &$+0.176$& BWL   &   2.49 &  4.70         \\         
361.7786 & 3.02 &$-0.029$& BWL+BK&   0.90 &  4.52         \\         
361.8768 & 0.99 &$-0.003$& BWL   &   7.16 &  4.52         \\         
362.2003 & 2.76 &$-0.150$& BWL   &   1.03 &  4.48         \\         
362.3186 & 2.40 &$-0.767$& BWL   &   0.58 &  4.50         \\         
363.8296 & 2.76 &$-0.375$& BWL   &   0.71 &  4.52         \\         
364.0389 & 2.73 &$-0.107$& BWL   &   1.26 &  4.51         \\         
380.5343 & 3.30 &$+0.312$& BWL   &   1.39 &  4.48         \\         
380.6696 & 3.27 &$+0.017$& BWL   &   0.96 &  4.56         \\         
380.7537 & 2.22 &$-0.992$& BWL   &   0.69 &  4.49         \\         
381.5840 & 1.49 &$+0.237$& BWL   &   7.99 &  4.57         \\         
382.0425 & 0.86 &$+0.119$& FMW   &   9.79 &  4.63         \\         
382.1178 & 3.27 &$+0.198$& BWL   &   1.38 &  4.56         \\         
382.5881 & 0.92 &$-0.037$& FMW   &   8.29 &  4.43         \\         
382.7823 & 1.56 &$+0.062$& FMW   &   6.14 &  4.32         \\         
384.0438 & 0.99 &$-0.506$& FMW   &   6.08 &  4.36         \\         
384.3257 & 3.05 &$-0.241$& BWL   &   0.72 &  4.48         \\         
384.9967 & 1.01 &$-0.871$& FMW   &   5.27 &  4.56         \\         
385.0818 & 0.99 &$-1.734$& FMW   &   1.73 &  4.59         \\         
385.2573 & 2.18 &$-1.185$& BWL   &   0.54 &  4.53         \\         
385.6372 & 0.05 &$-1.286$& FMW   &   7.53 &  4.70         \\         
385.9213 & 2.40 &$-0.749$& BWL   &   1.00 &  4.58         \\         
385.9911 & 0.00 &$-0.710$& FMW   &   9.83 &  4.75         \\         
386.5523 & 1.01 &$-0.982$& FMW   &   5.05 &  4.62         \\         
386.7216 & 3.02 &$-0.451$& BWL   &   0.65 &  4.62         \\         
387.2501 & 0.99 &$-0.928$& FMW   &   4.99 &  4.53         \\         
387.8018 & 0.96 &$-0.914$& FMW   &   5.31 &  4.56         \\         
389.5656 & 0.11 &$-1.670$& FMW   &   4.66 &  4.40         \\         
389.9707 & 0.09 &$-1.531$& FMW   &   5.81 &  4.50         \\         
390.2946 & 1.56 &$-0.466$& FMW   &   4.69 &  4.51         \\         
390.6480 & 0.11 &$-2.243$& FMW   &   2.59 &  4.53         \\         
391.7181 & 0.99 &$-2.155$& FMW   &   0.89 &  4.66         \\         
392.0258 & 0.12 &$-1.746$& FMW   &   5.29 &  4.62         \\         
392.7920 & 0.11 &$-1.522$& BWL   &   6.34 &  4.64         \\         
394.0878 & 0.96 &$-2.600$& FMW   &   0.29 &  4.56         \\         
395.6677 & 2.69 &$-0.429$& BWL   &   1.11 &  4.56         \\         
399.7392 & 2.73 &$-0.479$& BWL   &   1.16 &  4.66         \\         
400.5242 & 1.56 &$-0.610$& FMW   &   4.55 &  4.61         \\         
400.9713 & 2.22 &$-1.252$& BWL   &   0.47 &  4.55         \\         
401.4531 & 3.05 &$-0.587$& BWL   &   0.47 &  4.62         \\         
402.1867 & 2.76 &$-0.729$& BWL   &   0.55 &  4.58         \\         
404.5812 & 1.49 &$+0.280$& FMW   &   8.03 &  4.49         \\         
406.3594 & 1.56 &$+0.062$& BWL   &   7.19 &  4.55         \\         
407.1738 & 1.61 &$-0.022$& FMW   &   6.56 &  4.51         \\         
413.2058 & 1.61 &$-0.675$& BWL   &   3.99 &  4.59         \\         
413.4678 & 2.83 &$-0.649$& BWL   &   0.80 &  4.73         \\         
414.3415 & 3.05 &$-0.204$& BWL   &   1.17 &  4.66         \\         
414.3868 & 1.56 &$-0.511$& BWL   &   5.12 &  4.62         \\         
418.1755 & 2.83 &$-0.371$& BWL   &   1.43 &  4.74         \\         
418.7039 & 2.45 &$-0.548$& FMW   &   1.50 &  4.60         \\         
418.7795 & 2.42 &$-0.554$& FMW   &   1.49 &  4.58         \\         
419.1431 & 2.47 &$-0.666$& BWL   &   1.05 &  4.55         \\         
419.8304 & 2.40 &$-0.719$& FMW   &   1.64 &  4.77         \\         
419.9095 & 3.05 &$+0.155$& BWL   &   2.29 &  4.66         \\         
420.2029 & 1.49 &$-0.708$& FMW   &   4.59 &  4.63         \\         
421.0344 & 2.48 &$-0.928$& BWL   &   0.68 &  4.62         \\         
421.6184 & 0.00 &$-3.356$& FMW   &   0.55 &  4.71         \\         
421.9360 & 3.57 &$+0.000$& BWL   &   0.61 &  4.60         \\         
422.2213 & 2.45 &$-0.967$& FMW   &   0.65 &  4.61         \\         
422.7427 & 3.33 &$+0.266$& BWL   &   1.74 &  4.64         \\         
423.3603 & 2.48 &$-0.604$& FMW   &   1.40 &  4.64         \\         
423.5937 & 2.42 &$-0.341$& FMW   &   2.13 &  4.56         \\         
425.0119 & 2.47 &$-0.405$& FMW   &   1.93 &  4.60         \\         
425.0787 & 1.56 &$-0.714$& BWL   &   4.01 &  4.58         \\         
426.0474 & 2.40 &$+0.109$& BWL+BK&   4.12 &  4.52         \\         
427.1154 & 2.45 &$-0.349$& FMW   &   2.15 &  4.59         \\         
427.1761 & 1.49 &$-0.164$& FMW   &   7.09 &  4.65         \\         
428.2403 & 2.18 &$-0.779$& BWL   &   1.60 &  4.62         \\         
429.9235 & 2.42 &$-0.405$& BWL   &   2.14 &  4.62         \\         
432.5762 & 1.61 &$+0.006$& BWL   &   6.97 &  4.56         \\         
435.2735 & 2.22 &$-1.287$& BWL   &   0.53 &  4.62         \\         
437.5930 & 0.00 &$-3.031$& FMW   &   1.03 &  4.67         \\         
438.3545 & 1.49 &$+0.200$& FMW   &   8.64 &  4.68         \\         
440.4750 & 1.56 &$-0.142$& FMW   &   6.44 &  4.52         \\         
441.5123 & 1.61 &$-0.615$& FMW   &   4.36 &  4.59         \\         
442.7310 & 0.05 &$-2.924$& BWL   &   1.02 &  4.61         \\         
444.2339 & 2.20 &$-1.255$& FMW   &   0.54 &  4.57         \\         
445.9118 & 2.18 &$-1.279$& FMW   &   0.57 &  4.60         \\         
446.1653 & 0.09 &$-3.210$& FMW   &   0.62 &  4.68         \\         
448.2170 & 0.11 &$-3.501$& FMW   &   0.48 &  4.87         \\         
449.4563 & 2.20 &$-1.136$& FMW   &   0.73 &  4.59         \\         
452.8614 & 2.18 &$-0.822$& FMW   &   1.42 &  4.58         \\         
487.1318 & 2.87 &$-0.363$& BWL   &   1.02 &  4.55         \\         
487.2138 & 2.88 &$-0.567$& BWL   &   0.67 &  4.57         \\         
489.0755 & 2.88 &$-0.394$& BWL   &   0.97 &  4.56         \\         
489.1492 & 2.85 &$-0.112$& BWL   &   1.68 &  4.54         \\         
491.8994 & 2.87 &$-0.342$& BWL   &   1.13 &  4.58         \\         
492.0503 & 2.83 &$+0.068$& BWL   &   2.37 &  4.53         \\         
495.7597 & 2.81 &$+0.233$& BWL   &   3.27 &  4.55         \\         
\ion{Fe}{ii}                              \\         
423.3172 & 2.58 &$-1.900$& av    &   2.06 &  4.65         \\         
492.3927 & 2.89 &$-1.320$& av    &   3.21 &  4.58         \\         
\hline\hline\\
\end{longtable}
}

\end{document}